\RequirePackage{lineno} 

\documentclass[twocolumn,preprintnumbers,amsmath,amssymb,nofootinbib]{revtex4}

\usepackage{graphicx}
\usepackage{dcolumn}
\usepackage{bm}
 
\usepackage{epstopdf}
\def\bea{\begin{eqnarray}}
\def\eea{\end{eqnarray}}

\def\pp{\mbox{$p$-$p$}}

\def\pa{\mbox{$p$-A}}

\def\auau{\mbox{Au-Au}}

\def\pbpb{\mbox{Pb-Pb}}
\def\ppb{\mbox{$p$-Pb}}

\def\aa{\mbox{A-A}}
\def\nn{\mbox{N-N}}

\def\ep{\mbox{e-p}}
\def\ee{\mbox{$e^+$-$e^-$}}
\def\ppbar{\mbox{$p$-$\bar p$}}

\def\pt{$p_t$}
\def\mt{$m_t$}
\def\yt{$y_t$}

\def\nch{$n_{ch}$}
\def\mmpt{$\bar p_t$}
\def\k0s{$\text{K}^0_\text{S}$}

\begin{document} 

\setlength{\pdfpagewidth}{8.5in}
\setlength{\pdfpageheight}{11in}

\setpagewiselinenumbers
\modulolinenumbers[5]

\preprint{version 1.6}

\title{Few-gluon interactions and multipole radiation in high-energy nuclear collisions
}

\author{Thomas A.\ Trainor}\affiliation{University of Washington, Seattle, WA 98195}


\date{\today}

\begin{abstract}

Broad claims have been made over years about achievement of quark-gluon plasma (QGP) formation in high-energy heavy-ion collisions based on appearance of certain phenomena anticipated for QGP formation. More recently, similar phenomena have appeared in smaller collision systems where they were unexpected. In response, the original narrative associated with QGP formation has been altered to accommodate more-recent results, with introduction of novel concepts such as ``QGP droplets'' appearing even in \pp\ collisions. In contrast, alternative research has revealed novel aspects of \pp\ and \ppb\ collisions such as exclusivity for \nn\ interactions and consequences of time dilation for interacting partons. Collision geometry for A-B collisions has also been shifted from conventional Glauber Monte Carlo simulations (strongly biased) to inversion of ensemble-mean \mmpt\ data.
 The present study demonstrates that jet production dominates all aspects of $p_t$ spectrum structure and minimum-bias angular correlations over the full $p_t$ range of accessible data. Recently, progress has been made on interpretation of azimuth quadrupole ($v_2$) data, reexpressed in terms of {\em total correlated-pair number} as an {\em extensive} measure, leading to inference of quadrupole $m_t$ spectra and quadrupole amplitude variation across all A-B collision systems that show strong indication of the effects of exclusivity. The same approach applied to jet angular correlations shows similar trends. A comprehensive quantitative description of the two QCD phenomena across all collision systems has emerged. The underlying processes are few-gluon interactions producing characteristic correlation structures corresponding to color-dipole (two-gluon, dijet) and color-quadrupole (three-gluon) radiation. That description does not rely on any role for a dense medium, multiple scattering, QGP droplets  or hydro theory. It applies the same rules uniformly to small and large collision systems.

\end{abstract}   

\maketitle

 \section{Introduction}

The intended goal of the relativistic heavy-ion collider (RHIC) had been production of a conjectured quark-gluon plasma (QGP) and study of its properties. Motivation for the RHIC consisted in part of extrapolating results from the Bevalac program (c.\ 1980), including the discovery of nucleon plastic flow (``collectivity'') in \mbox{A-B} collisions, to much higher collision energies and particle densities where partonic (quark and gluon) degrees of freedom were expected to dominate within  high-density and high-temperature A-B collision systems. In preparation for the RHIC program  various proposals for experimental verification of QGP formation were presented~\cite{harris}.

Extrapolating nucleon angular distributions from the Bevalac to much higher energies led to an argument via angular correlation measure $v_2$ that has been associated with {\em elliptic flow}, a variant of the nucleon flow observed at the Bevalac: ``We show that anisotropies in transverse-momentum distributions provide an {\em unambiguous signature} of transverse collective flow in ultrarelativistic nucleus-nucleus collisions [emphasis added]''~\cite{ollitrault}. Collective flow was then seen as a primary verification of QGP formation. $v_2$ data have since played a central role in  claims of achieving QGP formation~\cite{perfect,qgp1,qgp2,keystone}. 

A related argument referring to other aspects of collision data has been elaborated. In contrast to nucleon plastic flow at Bevalac energies, \aa\ collisions at much higher energies were expected to include copious jet production based on results from high-energy physics experiments at the S$ {p \bar p}$S, Fermilab, LEP and HERA. A narrative developed for the RHIC program (e.g.\ Ref.~\cite{starspec04}) partitioned the  transverse-momentum \pt\ spectrum into a low-\pt\ interval including ``soft'' physics ($< 2$-3 GeV/c), a high-\pt\ interval including ``hard'' or jet physics ($> 5$-6 GeV/c) and an intermediate region subject to continuing speculation. The interval said to manifest jet production might relate to ``jet quenching'' (reduction of jet fragments via attenuation) as evidence for a  dense QGP. In general, preferred terminology, analysis methods and plotting formats have tended to favor a desired outcome.

White papers produced by the four RHIC collaborations in 2004 summarized evidence interpreted to support QGP formation and concluded that formation was confirmed by \auau\ data~\cite{starwhite}. Based in part on hydrodynamic (hydro) theory descriptions of $v_2$ data and in part on interpretation of spectrum ratio $R_\text{AA}$ as providing evidence for jet quenching, it was then concluded that \auau\ collisions at RHIC produced a ``perfect fluid''~\cite{perfect}.

However, with commencement of LHC operations in 2010 experimental evidence emerged that certain data characteristics, attributed to QGP formation in RHIC \auau\ collisions, were also observed in ``small systems'' --  \ppb\ and even \pp\ collisions~\cite{cmsridge}. The preferred response was to revise the conventional narrative to favor QGP formation in smaller systems, even \pp\ collisions, in terms of ``QGP droplets'' \cite{buzsa}. Such revisions persist to the present,  e.g.\ as reported for 200 GeV $d$-Au collisions~\cite{phenixdroplets}.

An alternative strategy might have been to reexamine the legitimacy of {\em any} data interpretations favoring QGP formation. The present study summarizes efforts in that direction and presents an alternative narrative based on fundamental QCD processes (elementary partonic collisions) requiring updated interpretations based on experience gained over twenty-five years. This study considers recent evidence for {\em few-gluon dominance} of all aspects of A-B collisions and apparent consequences of relativistic and quantum effects within A-B collisions.

This article begins with a brief summary of general assertions that a QGP had been achieved at the RHIC based on certain phenomena argued to signal QGP formation in \aa\ collisions and further arguments thereafter that unexpected appearance of those same phenomena in smaller collision systems at the LHC also indicated QGP formation therein. 
The article then proceeds to demonstrate the dominant role of jet production in hadron yields and \pt\ spectra despite a general tendency to minimize recognition of jet production in favor of a monolithic ``flow-centric'' model of collision dynamics.

The article then shifts to phenomena observed in two-particle angular correlations, focusing on both jet-related and quadrupole-related features, where ``quadrupole'' here refers to a correlation feature conventionally associated with the $v_2$ measure. Such studies have lead recently to the discovery that the quadrupole amplitude as described by an {\em extensive} measure follows a trend {\em predicted} on the basis of quadrupole measurements for \pp\ collisions and a new method for determining A-B collision geometries over all collision systems from \pp\ to central \aa. The same method applied to minimum-bias (no \pt\ cuts) jet-related correlations arrives at a corresponding predicted trend for dijets with similar results.

One may conclude in general that hadron production in A-B collisions arises from three processes: (a) dissociation of participant nucleons to form a soft component, (b) two-gluon interactions generating color-dipole radiation (dijets) and (c) {\em three}-gluon interactions generating color-quadrupole radiation. Individual interactions appear to be independent and observed correlation patterns appear to be linear superpositions. An A-B collision process may then be described as few-gluon interactions resulting in QCD multipole radiation. There seems to be no requirement for a dense medium or a hydrodynamic description.

This article is arranged as follows:
Section~\ref{promise}  reviews claims for successful QGP production at the RHIC and followup revisions of criteria to accommodate similar results in smaller systems at the LHC.
Section~\ref{structure} demonstrates the dominant effects of jet production in \pt\ spectra for all accessible data.
Section~\ref{heavyspectra}  examines detailed structure of \pt\ spectra for heavy hadrons and demonstrates that for $\Xi$s and $\Omega$s almost all detected particles are jet fragments.
Section~\ref{mptsec}  demonstrates that ensemble-mean \mmpt\ trends correspond precisely to properties of spectrum hard components representing jet fragments.
Section~\ref{enhance}  suggests that for any significant variation in hadron fractional abundances (e.g.\ relating to ``strangeness enhancement'') jet production cannot be ruled out as the source.
Section~\ref{2dcorr}  examines results from 2D angular correlations and reveals that correlation amplitudes measured by number of correlated pairs follow {\em predicted} trends based on \pp\ measurements and collision geometries  derived from ensemble-mean \mmpt\ data and exclusivity of \nn\ collisions.
Sections~\ref{disc} and~\ref{summ} present discussion and summary. 
Appendix~\ref{tcmpid} defines a TCM for identified-hadron (PID) A-B spectra
 and Appendix~\ref{geometry} describes a geometry model for A-B collisions.

\section{Broad claims for QGP formation} \label{promise}

Since the startup of the RHIC in 2000 several published review articles addressing the question of QGP formation in high-energy collisions have received broad attention. In this section three such papers are reviewed briefly as providing a context for the analysis material presented in this study. They are later reconsidered in the discussion section in light of those analysis results. The first paper considers evidence for a quark-gluon plasma in \auau\ collisions from four RHIC experiments. The second and third papers review claimed evidence for formation of ``QGP droplets'' within smaller collision systems as produced at the LHC. References in this section include publication dates and number of citations to date.

\subsection{Early RHIC interpretation: perfect fluid} \label{perfect}

Reference~\cite{perfect} (2004, 1431) reviews experimental evidence reported in ``white papers'' presented by the four RHIC collaborations after nearly four years of operation.

Commenting in its introduction on the body of RHIC experimental results up to mid 2004 Ref.~\cite{perfect} notes  ``...the data are so striking and decisive that several strong physics conclusions can already be drawn. They [data] establish empirically that
a special form of strongly coupled QGP (sQGP) exists with remarkable properties.''

``In heavy ion reactions the flow pattern of thousands of produced hadrons is the primary observable used to look for novel collective phenomena. The collective flow properties test two of the conditions necessary for the validity of the QGP hypothesis.'' Those conditions are the extent of thermalization and the equation of state.

``Elliptic flow measurements confirm that the quark-gluon matter produced at RHIC is to a very good approximation in local thermal equilibrium up to about 3 fm/c. In addition, the final hadron mass dependence of the flow pattern is remarkably consistent with numerical QCD computations of the equation of state. Viscous corrections furthermore appear to be surprisingly small during this early evolution. The produced Quark Gluon Plasma must therefore be very strongly interacting.''

``Theoretical analysis of jet quenching confirm [sic] the energy density estimates determined from measurements of particle multiplicity. They give large energy losses for jets propagating through the matter produced at RHIC, and strengthen the case for {\em multiple strong interactions} [emphasis added] of the quark and gluon constituents of the matter made at RHIC.''

``Our criteria for the discovery of the Quark Gluon Plasma at RHIC are:
(a) Matter at energy densities so large that the simple degrees of freedom are quarks and gluons. This energy density is that predicted by lattice gauge theory [LGT] for the existence of a QGP....
(b) The matter must be to a good approximation thermalized.
(c) The properties of the matter [...]\ while it is hot and
dense must follow from QCD computations based on hydrodynamics, lattice
gauge theory results, and perturbative QCD for hard processes such as jets.
All of the above are satisfied from the published data at RHIC.''

Such interpretations are strongly dependent on $v_2$ data as indicating large-scale collective flow of a dense medium (that is, affecting a large majority of produced particles). Comparisons with hydro theory then lead to strong conclusions. Interpretation is further dependent on identifying spectrum evolution at higher \pt\ as indicating ``jet quenching'' wherein energetic high-\pt\ partons encounter a dense QCD medium and lose substantial energy. If  conventional interpretations of $v_2$ data are invalid and high-\pt\ spectrum evolution is differently interpreted in an alternative scenario  such claims may be falsified. 

\subsection{Dense medium in small systems I} \label{small1}

Reference~\cite{buzsa} (2018, 1029) emphasizes recent observations that phenomena appearing in more-central \aa\ collisions and interpreted as QGP manifestations appear also to varying extent in \pa\ and even \pp\ collisions.

Regarding jet production ``A small fraction of the incident partons suffer [sic] hard perturbative interactions
as the discs overlap initially which, as we will discuss later, lead to a relatively improbable but very important production of particles with high transverse momentum.''

``Heavy ion collisions quickly form a droplet of quark-gluon plasma
(QGP) with a remarkably small viscosity.'' Such an unlikely scenario requires a proposed mechanism and acknowledgement of any evidence to the contrary.

``They [partons] are so strongly coupled to each other that they form a collective medium that expands and flows as a relativistic hydrodynamic fluid with a remarkably low viscosity to entropy density ratio $\eta/s \approx 1/4\pi$, in units with $\hbar = k_B = 1$, within a time that can be shorter than or of order 1 fm/c in the rest frame of the fluid.''

``What did come as a surprise is how many other phenomena are similar in AA and pA collisions, and even in pp collisions, in particular when the comparison is done between collisions in different systems with the same final state particle density dN/dy. Examples include the rapidity distribution, particle spectra, particle ratios including those involving strangeness, and, most significantly, the azimuthal anisotropies  encoded in multiparticle correlations that were once thought to be unique to AA collisions.''

``...it is tempting to interpret them [similarities among pp, pA and AA] as indicating that proton-sized droplets of QGP can be formed in those pp and pA collisions that produce final states with sufficiently large dN/dy.''

``{In occasional heavy ion collisions, partons from the incident nuclei scatter off each other at very large momentum transfer, creating two or more quarks or gauge bosons [gluons?] with very high transverse momentum (many tens of GeV at RHIC; as high as 100 or even 1000 GeV at the LHC)}. When such a hard scattering occurs in a proton-proton collision, each hard parton that is produced showers into a spray of softer partons within some irregular cone in momentum space, called a jet. Jet production and showering in vacuum is well described by perturbative QCD. When a jet is produced in a heavy ion collision, the partons in the shower must plow through the droplet of QGP produced in the same collision. As this happens, the jet partons: (i) lose energy and forward momentum, (ii) pick
up momentum transverse to their original direction, and (iii) deposit energy and momentum into the droplet of QGP, creating a wake.'' 

In that narrative jets are assumed to be rare and restricted to high parton energy and high hadron \pt; all else is relegated to ``soft'' physics -- QGP and flows. Most of the jet energy spectrum (down to 3 GeV per UA1 observations~\cite{ua1jets}) is excluded from the preferred narrative of a ``proton-sized'' QGP droplet~\cite{phendroplets,phenixdroplets} imposed {\em a priori}. The tiny fraction of jets acknowledged must ``plow through the droplet'' and be attenuated (quenched). 

\subsection{Dense medium in small systems II} \label{small2}

Reference~\cite{nagle} (2018, 433) also confronts appearance of QGP-like manifestations in small collision systems.

``Crucial information regarding collectivity is garnered through the measurement of two or more particle correlations, often parameterized via the particles' relative azimuthal angle [$\phi_\Delta = \phi_1 - \phi_2$] in the transverse plane, and their relative longitudinal pseudorapidity [$\eta_\Delta = \eta_1 - \eta_2$].''

2D angular correlations on $(\eta_\Delta,\phi_\Delta)$ do present essential information but must be quantified and interpreted carefully. Correlation data should be processed as intact 2D distributions. Model fits to 2D data permit accurate separation of contributions from several mechanisms.

``... these dijet correlations are subdominant for $p_t < 5$ GeV/c.'' The term ``subdominant'' is by convention a musical term having to do with chord structure. Here the term appears as a neologism suggesting that dijet correlations are relatively negligible below 5 GeV/c.

``It now may seem surprising that no jet quenching effect is apparent in p + A collisions if indeed a hot medium is formed.''

``In summary, ample theoretical arguments developed over the past decade suggest that viscous relativistic hydrodynamics can be applied to describe particle production and flow in p + p and p + A collisions at high energies.''

``As of early 2018, the field of relativistic heavy ion physics is in the midst of a revolution in our understanding of the conditions necessary for nuclear matter to behave as a near-perfect fluid with bulk dynamics described by viscous relativistic hydrodynamics. ... The revolution has been driven by the experimental observation of {\em flow-like features} [emphasis added] in the collisions of small hadronic systems''

Such arguments require that once conjectured characteristics of sought-after QGP production are identified in A-A collisions their unexpected appearance in small systems must continue  to be associated with QGP production no matter how unlikely that might seem in the context of high-energy physics findings.
But the scientific method {\em requires} that small systems, once assigned as references, should not be reinterpreted in terms of a scenario developed in the past for QGP claims in \aa\ collisions. They should  be fully understood {\em ab initio}  and then employed to {\em reinterpret} \aa\ data. What follows is a summary of recent developments along those lines.

\section{TCM Spectrum structure $\bf vs$ data} \label{structure}

In Ref.~\cite{ppprd} it is demonstrated that a two-component (soft+hard) spectrum model (TCM) as presented in Eq.~(\ref{tcm}) is {\em required} by  200 GeV \pp\ \pt\ spectra. In this section that argument is applied to 13 TeV \pp\ data~\cite{alippss} and 5 TeV \ppb\ data~\cite{aliceppbpid}. 
Spectra for neutral \k0s\ are preferred because V0 detection extends  particle \pt\ acceptance down to zero.
The TCM applied to {\em identified-hadron} (PID) spectra for A-B collisions requires accurate determination of A-B collision geometry per App.~\ref{geometry}.

The TCM is inferred from spectrum data without  {\em a priori} assumptions.  It typically describes data within point-point uncertainties.  Applied over twenty years to numerous collision systems, it is an example of {\em lossless data compression}: redundant data elements are discarded while all significant information carried by data is retained.
The TCM generates {\em predictions} for given  manifestations (spectra, correlations) within data uncertainties. As a result, fundamental  degrees of freedom are revealed.
This section provides extensive evidence for the {\em dominant role of jet production} in spectrum evolution.

\subsection{PID spectra and running integrals} \label{}

$\bar \rho_{0}(p_t)$ spectra for  \k0s\   below are published densities on \pt\ presented here as $d^2n_{ch} / p_t dp_t dy_z$ plotted vs transverse rapidity \yt. Spectra are rescaled by soft-component densities $\bar \rho_{si}(n_{ch}) = z_{si}(n_{ch}) (N_{part}/2) \bar \rho_{sNN}$ 
as defined in App.~\ref{tcmpid} and \ppb\ $N_{part}$ trends reported in Ref.~\cite{ppbpid}. Rescaled spectra for  hadron species $i$ are denoted by
\bea \label{xi}
X_{i}(p_t) &\equiv& \bar \rho_{0i}(p_t) / \bar \rho_{si}
\\ \nonumber
&\approx& \hat S_{0i}(p_t) + \tilde z_i x \nu \hat H_{0i}(p_t),
\eea
in which case if  $\bar \rho_{si}$ factors are estimated accurately the spectra should coincide at low \pt\ where hard components (jets) don't contribute. Note that  $\bar \rho_{si}$ are here {\em predicted} from previous analysis, not fitted to individual spectra.

\begin{figure}[h]
	\includegraphics[width=3.3in]{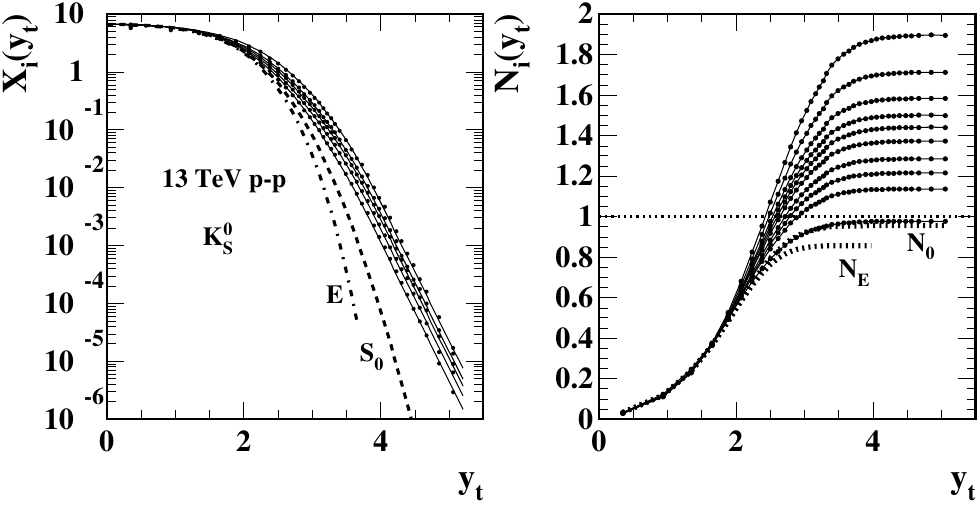}
\put(-140,105) {\bf (a)}
\put(-57,105) {\bf (b)}\\
	\includegraphics[width=3.3in]{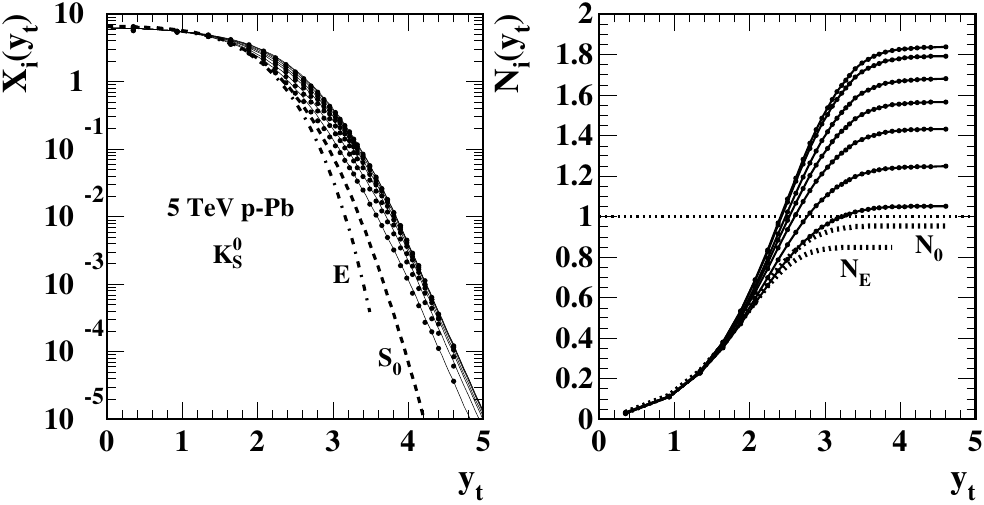}
\put(-140,105) {\bf (c)}
\put(-57,105) {\bf (d)}
	\caption{\label{kosspectra}
Left: Rescaled \k0s\ spectra as defined by Eq.~(\ref{xi}) for 13 TeV \pp\ collisions (every other \pp\ event class for visibility) (a) and 5 TeV \ppb\ collisions (c). Solid curves are corresponding TCMs.
Right: Running integrals as defined in Eq.~(\ref{runint}) for data in left panels. $N_0(p_t)$ are running integrals of soft components $\hat S_0(p_t)$. $N_E(p_t)$ are running integrals of corresponding exponentials $E$. Curves simply connect the points.
}  
\end{figure}

Figure~\ref{kosspectra} (left) shows rescaled spectra $X_i(y_t)$ as in Eq.~(\ref{xi}) plotted vs transverse rapidity $y_{t} = \ln[(p_t + m_{t})/m]$ (for $\pi$s) that greatly improves visual access to lower-\pt\ structure. Note that $y_t = 1$ is $p_t \approx 0.15$ GeV/c, $y_t = 2$ is $p_t \approx 0.5$ GeV/c, $y_t = 2.7$ is $p_t \approx 1$ GeV/c and $y_t = 4$ is $p_t \approx 4$ GeV/c. Data are shown as solid dots, and solid curves are the TCM from Ref.~\cite{ppsss}. The dashed curves are soft-component models $\hat S_0(y_t)$ from Refs.~\cite{alippss,aliceppbpid} -- Boltzmann exponentials on transverse mass $m_{ti}$ (for kaons) with power-law tails. Dash-dotted curves E are pure exponential distributions arising if the power-law exponent $n$ for $\hat S_0(y_t)$ goes to the limit $1/n \rightarrow 0$.  The $\hat S_0(p_t)$ normalization rescale is retained for E as well.

Figure~\ref{kosspectra} (right) shows running integrals of the form
\bea \label{runint}
N_{i}(p_t) &\equiv& \int_0^{p_t} dp_t' p_t' X_{i}(p_t')
\\ \nonumber
&\rightarrow& 1 + \tilde z_i x \nu ~~~\text{for $p_t \rightarrow \infty$}.
\eea
where the second line follows from Eq.~(\ref{xi}) with $\tilde z_i x\nu =  \bar\rho_{hi}/ \bar \rho_{si}$.
If $X_{i} \rightarrow \hat S_{0i}$ then $N_{i} \rightarrow N_0$ which should  go asymptotically to 1. However, data integrals are discrete on data \pt\ values, so $N_0(y_t)$ terminates a few percent below 1 whereas integrated $\hat S_{0i}$ on continuum \pt\ do go to 1. Note that lower-\pt\ parts of  running integrals coincide across event classes within point-point uncertainties demonstrating that soft components are insensitive to dynamics occurring {\em within} the collision space-time volume.

The plotting format in Fig.~\ref{kosspectra} (left) is desirable for two reasons: 
(a) Soft component $\hat S_0(y_t)$ that describes low-\pt\ data within their statistical uncertainties for \ppb\ and \pp\ collisions~\cite{pidpart1,pidpart2} follows a Boltzmann exponential on \mt\ at lower \mt.
(b) Hard component $\hat H_0(y_t)$ that describes high-\pt\ data within data uncertainties for \ppb\ and \pp\ collisions follows an exponential $\exp(-q\, y_t)$ at higher \yt\ equivalent to a power law on \mt\ or \pt\ that manifests in this format as a straight line for $y_t > 4$. 

\begin{figure}[h]
	\includegraphics[width=1.65in]{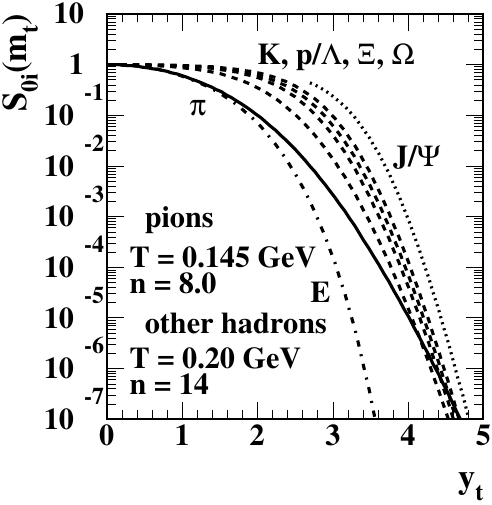}
	\includegraphics[width=1.65in]{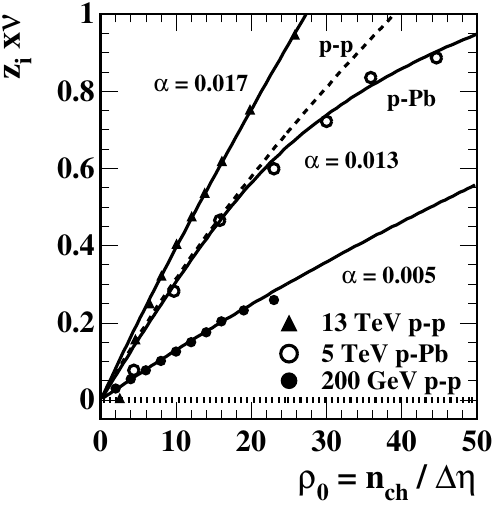}
	\caption{\label{endpoints}
		Left: Soft-component model $S_{0i}(m_t)$ evolution with hadron mass from $\pi$s to J/$\psi$s. The functions are scaled to pass through (0,1) for comparison. Those trends do not vary with specific A-B collision systems. Aside from the parameters indicated on the plot the shape evolution is determined solely by hadron mass.
		Right: Running integral endpoint data (points) from Fig.~\ref{kosspectra} (right). Curves are $\tilde z_i(n_{ch}) x \nu$ trends with $\tilde z_i(n_{ch})$ following the kaon straight line in Fig.~\ref{tildezparamss} (left).
	}  
\end{figure}

Fig.~\ref{endpoints} (left) shows soft components $\hat S_{0i}(m_{ti})$ (curves) for several hadron species from $\pi$s to J/$\psi$s. The comparison is meant to demonstrate that $\hat S_{0i}(m_{ti})$ are not fitted to specific data, instead reflect a general trend among hadron spectra. Slope parameter $T$ is constrained by low-\pt\ data and predicted $\bar \rho_{si}$. That combination plus the unit-normal condition for $\hat S_{0i}$ then constrains power-law exponent $n$. That observation is supported by exponential curve E (dash-dotted). Function E is the asymptotic limit of $\hat S_{0i}$ if $1/n \rightarrow 0$. Its running integral $N_\text{E}$ terminates well below unity because E lacks the exponential tail of $\hat S_{0i}$ required by spectrum data. As in previous studies~\cite{ppprd,ppbpid,pidpart1,ppsss}, $\hat S_{0i}$ is {\em defined} as the asymptotic limit of rescaled data spectra $X_{i}(p_t,n_{ch})$ as $n_{ch} \rightarrow 0$. The $\hat S_{0i}$ are approximately independent of collision system.

\subsection{Running-integral endpoints with predictions} \label{}

As noted for Eq.~(\ref{runint}) the asymptotic limit of a running integral in Fig.~\ref{kosspectra} (right) is $1 + \tilde z_i x\nu$. The values should be predicted from previous studies of \pp\ and \ppb\ systems. 

Figure~\ref{endpoints} (right) shows endpoint values (minus 1) from Fig.~\ref{kosspectra} (right) for LHC data (triangles, open circles) and from Fig.~7 (right) of Ref.~\cite{ppprd} for 200 GeV \pp\ data (solid dots). The points are derived from data integrals with no assumptions except for $\bar \rho_{si}$ derived from previous studies. For \k0s\ data $\tilde z_i \approx 2.7$ per Fig.~\ref{tildezparamss} (left). For nonPID 200 GeV \pp\ spectra fixed value $\tilde z_i = 2.7$ is included here as a multiplier for comparison with the LHC \k0s\ data.

The solid curves describing data are constructed as follows: For \pp\ spectra one expects $x \approx \alpha \bar \rho_s$ and $\nu = 1$. 
$\bar \rho_s$ is obtained from $\bar \rho_0$ by quadratic equation solution
\bea \label{barrhos}
\bar \rho_s &=& \frac{\sqrt{1 + 4 \alpha \bar \rho_0} - 1}{2 \alpha}
\eea
and $\alpha$ depends on collision energy~\cite{alicetomspec}.
For \ppb\ spectra product $x\nu$ is determined via analysis of \mmpt\ data~\cite{pppid}. Increase of $\tilde z_i(n_s)$ with event class as in Fig.~\ref{tildezparamss} (left) is required to describe the LHC PID endpoint data precisely. 

\subsection{Spectrum hard-component properties} \label{}

As noted in Ref.~\cite{ppprd} the running integrals in Fig.~\ref{kosspectra} (right) anticipate the general properties of spectrum hard components. Subtraction of the soft-component contribution $N_0(y_t)$ leaves trends well approximated by an error function (erf) which is the running integral of a Gaussian.
Spectrum hard components, rescaled by factor $\bar \rho_{si}$, are simply complements of soft-component model $\hat S_{0i}(p_t)$: $X_i(p_t) - \hat S_{0i}(p_t) \approx \tilde z_i x \nu \hat H_{0i}(p_t)$ per Eq.~(\ref{xi}). When transformed to densities on transverse rapidity \yt\ they have a Gaussian shape consistent with the running integrals.

\begin{figure}[h]
	\includegraphics[width=3.3in]{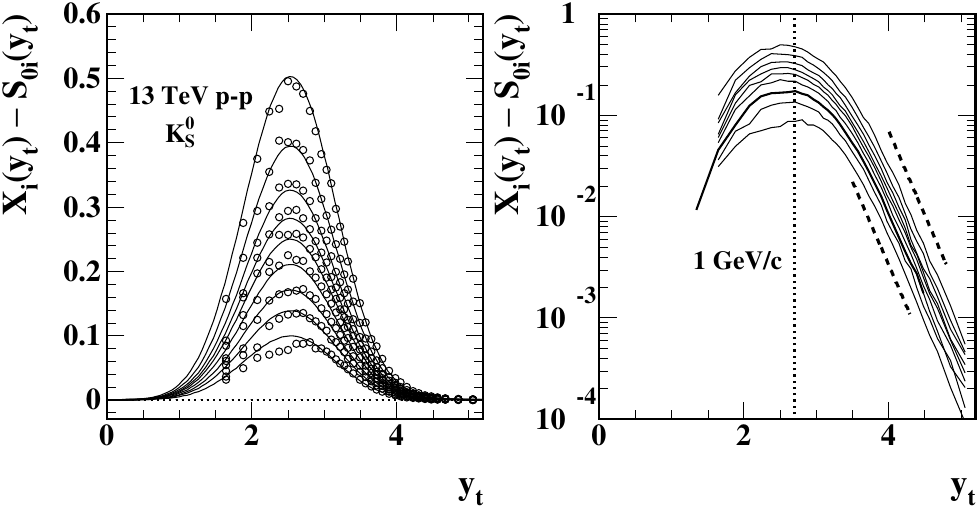}
\put(-140,105) {\bf (a)}
\put(-20,105) {\bf (b)}\\
\includegraphics[width=3.3in]{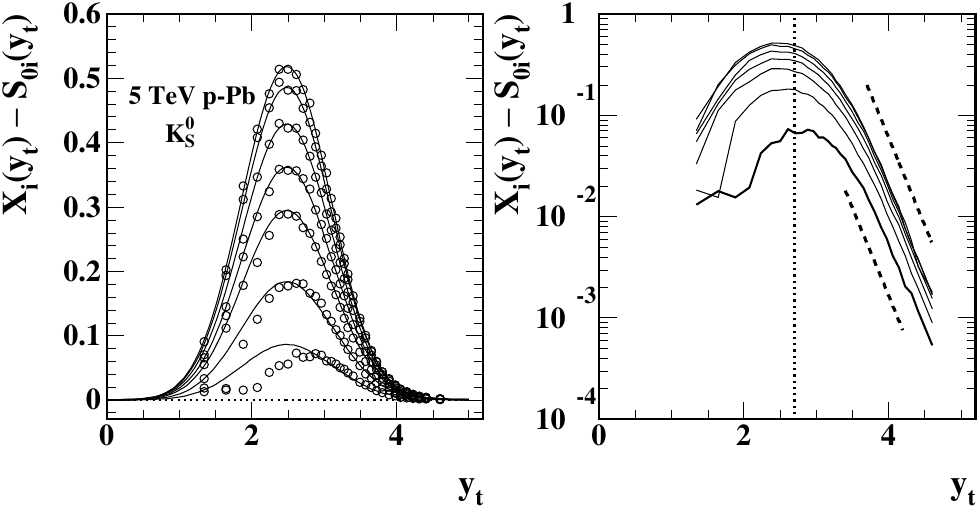}
\put(-140,105) {\bf (c)}
\put(-20,105) {\bf (d)}
	\caption{\label{piddataa}
Left: \k0s\ hard component data (points) of the form $X_i(p_t) - \hat S_{0i}(p_t)$ for 13 TeV \pp\ collisions (omitting the lowest-\nch\ event class) (a) and 5 TeV \ppb\ collisions (c) plotted in a linear format vs transverse rapidity \yt. Curves are corresponding TCM hard components~\cite{ppsss}.
Right: Hard-component {\em data} from the left panels (points) plotted as solid curves in a semilog format. Data at higher \pt\ indicate power-law trends (dashed lines) consistent with jet production.
} 
\end{figure}

Figure~\ref{piddataa} (left) shows differences $X_i(p_t) - \hat S_{0i}(p_t)$ transformed to densities on \yt\ via Jacobian factor $p_t m_{ti} / y_{ti}$. Spectrum data are represented by points whereas curves correspond to TCMs as shown in Fig.~\ref{kosspectra} (left). The  result is dependent only on factors $\bar \rho_{si}$ as defined in Eq.~(\ref{zxi}). On these linear plots the results are clearly well approximated by Gaussians on \yt\ which then suggests the correct model function {\em as required by data}.
For \pp\ event class 10 the hard component has a very small amplitude. The integral end point in Fig.~\ref{kosspectra} (b) is barely above the $N_0(y_t)$ soft reference (dotted), with similar indication in Fig.~\ref{endpoints} (right).  $\bar \rho_0$ for that event class is less than one-half of the 13 TeV NSD value $\approx 6.5$. Event-class 10 is therefore omitted from Fig.~\ref{piddataa} (a). Note that the \k0s\ hard component (jets)  persists down to $y_{t\pi} \approx 1.5$ ($p_t \approx 0.3$ GeV/c).

Figure~\ref{piddataa} (right) shows the same results on semilog plots. 
In this plotting format exponential (on \yt) tails are clearly evident, as characterized by the dashed lines. An exponential tail on \yt\ is equivalent to a power-law tail on \pt\ that relates to an underlying minimum-bias jet energy distribution~\cite{fragevo,jetspec2}. Exponents for hard-component power-law trends in panels (b,d) follow jet-related energy dependences reported in Ref.~\cite{jetspec2}.
High-\pt\ data trends exhibit regular spacing for \pp\ but {\em significant irregularities} for \ppb\ data. The difference is explained by a space-time argument in Ref.~\cite{tompbpb}. Such variation of \ppb\ spectra is not related to jet suppression in a dense medium.

Hard-component parameters for \k0s\ spectra reported in Rev.~\cite{ppsss} are as follows: for \pp\ the parameters are $\bar y_t = 2.7$, $\sigma_{y_t} = 0.58$ and $q = 3.7 \leftrightarrow\ n = 5.9$. For \ppb\ the parameters are $\bar y_t = 2.65$, $\sigma_{y_t} = 0.59$ and $q = 3.95 \leftrightarrow\ n = 6.15$, consistent with previous analyses over years and with QCD collinear factorization given a jet energy spectrum with effective lower bound near 3 GeV. As noted above, all aspects of spectrum {\em evolution} for these systems are determined by jet production.

\section{Heavy hadron spectra} \label{heavyspectra}

Jet production plays a dramatic role in the abundances of heavy hadrons as indicated by the trend $\tilde z_i \propto$ hadron mass $m_i$ in Fig.~\ref{tildezparamss} (right). In the previous section the major roll of jet production in {\em any} hadron spectrum was established for \k0s.
In this brief section the relation between jets and ``heavy hadrons'' is probed in terms of spectrum structure for four hadron species. The focus is on more-massive baryons, $\Xi$s and $\Omega$s. \k0s\ and $\Lambda$s are included as references. TCM analysis details are presented in Ref.~\cite{ppsss}. 
Small collision systems \pp\ and \ppb\ are compared where recent claims of QGP production have been made~\cite{buzsa,nagle}. For heavier hadrons {\em jet production dominates  measured spectra}. 

\begin{figure}[h]
	\includegraphics[width=1.62in]{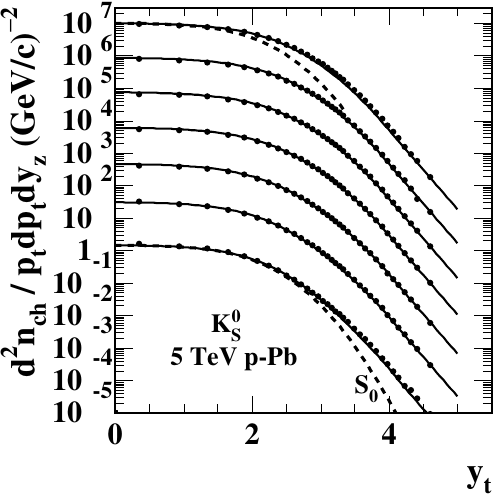}
	\includegraphics[width=1.65in]{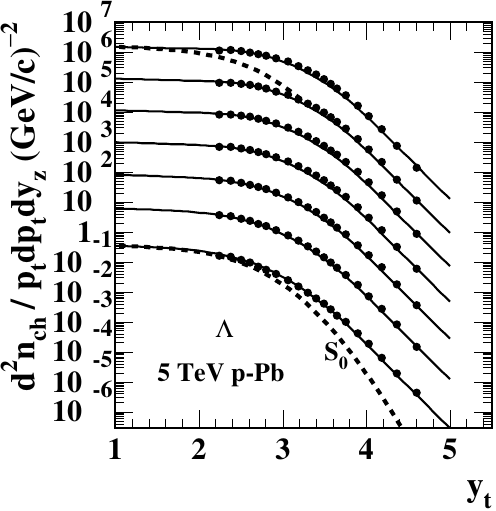}
	\put(-145,105) {\bf (a)}
	\put(-23,105) {\bf (b)}
	\\
	\includegraphics[width=1.65in]{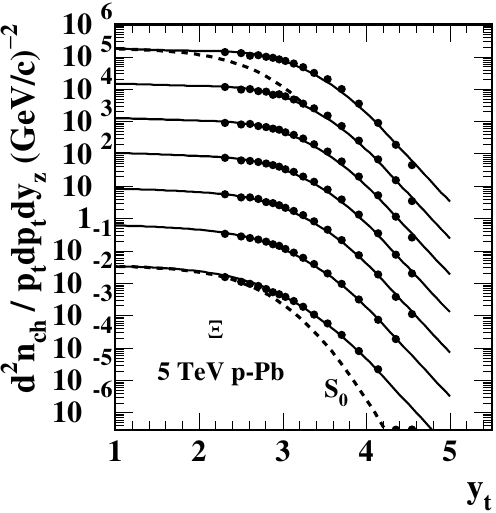}
	\includegraphics[width=1.65in]{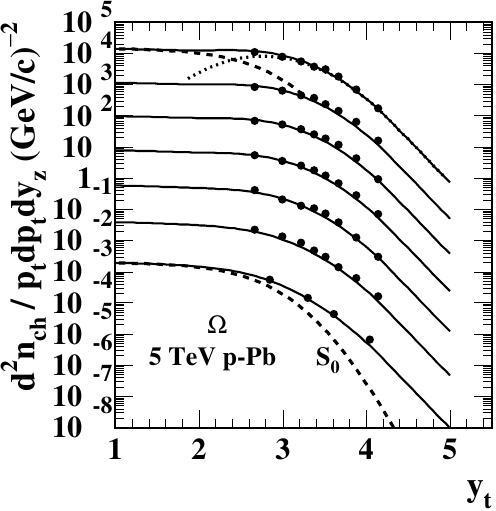}
	\put(-145,105) {\bf (c)}
	\put(-23,105) {\bf (d)}
	\\
	\caption{\label{ppbpiddata}
		\pt\ spectrum data (points) for strange hadrons from 5 TeV \ppb\ collisions:
		(a) \k0s and
		(b) $\Lambda$s from Ref.~\cite{aliceppbpid},
		(c) $\Xi$s and
		(d) $\Omega$s from Ref.~\cite{alippbss}.
		Solid curves represent the TCM. Dashed curves represent TCM soft components in the form $z_{si}(n_s) \bar \rho_s \hat S_0(p_t)$. The dotted curve in (d) is   the TCM hard component in the form $z_{hi}(n_s) \bar \rho_h \hat H_0(p_t)$. Spectra have been rescaled by powers of ten.
	} 
\end{figure}

Figure~\ref{ppbpiddata} shows 5 TeV \ppb\ PID \pt\ spectra (points) for (a) \k0s\, (b) $\Lambda$ (c) $\Xi$ and (d) $\Omega$ vs $\pi$ rapidity \yt. Solid curves are TCM predictions. TCM hard-component modes $\bar y_t$ vary with event class per Fig.~\ref{ybarcomp} (left). Dashed curves are soft-component models $z_{si}(n_s) \bar \rho_s \hat S_{0i}(p_t)$. The dotted curve in (d) is $\Omega$ hard-component model $z_{hi}(n_s) \bar \rho_h \hat H_{0i}(p_t)$. The last serves to demonstrate that {\em a large fraction of all $\Omega$s from central \ppb\ collisions are jet fragments.}

\begin{figure}[h]
	\includegraphics[width=1.65in]{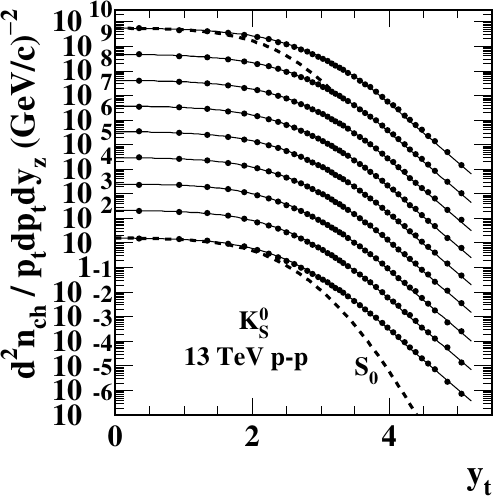}
	\includegraphics[width=1.65in]{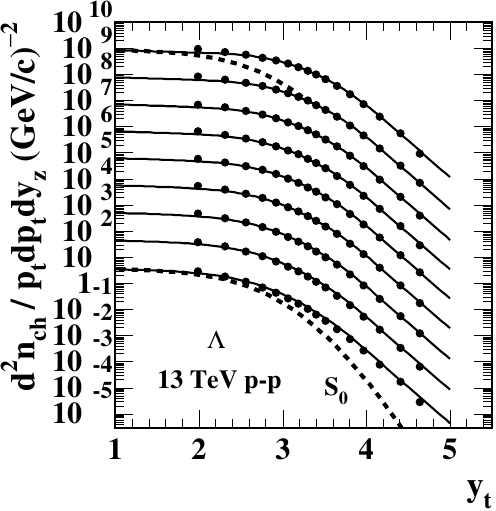}
	\put(-143,105) {\bf (a)}
	\put(-21,105) {\bf (b)}
	\\
	\includegraphics[width=1.65in]{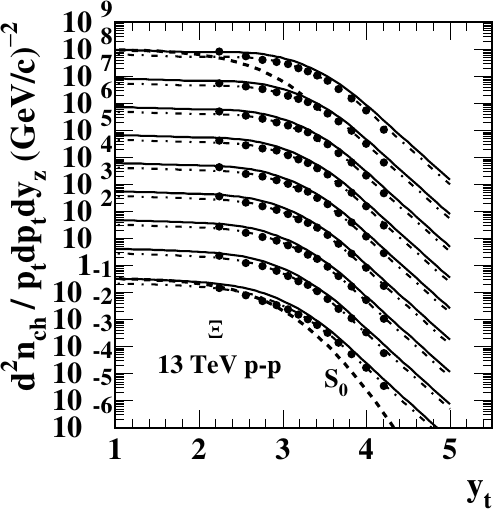}
	\includegraphics[width=1.65in]{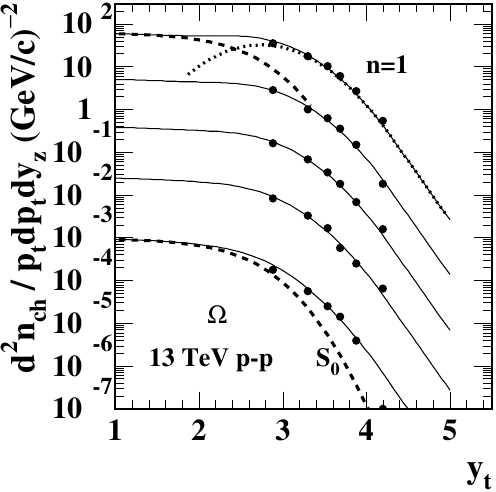}
	\put(-143,105) {\bf (c)}
	\put(-21,105) {\bf (d)}
	\\
	\caption{\label{piddata}
		\pt\ spectrum data (points) for strange hadrons:
		(a) \k0s,
		(b) $\Lambda$s,
		(c) $\Xi$s and
		(d) $\Omega$s from Ref.~\cite{alippss}.
		Solid curves represent the PID spectrum TCM. The dash-dotted curves in (c) are the TCM curves reduced by factor 2/3. The dashed curves are the TCM soft component in the form $z_{si}(n_s) \bar \rho_s \hat S_0(y_t)$.  The dotted curve in (d) is   the TCM hard component in the form $z_{hi}(n_s) \bar \rho_h \hat H_0(p_t)$.  Spectra have been rescaled by powers of ten.
	} 
\end{figure}

Figure~\ref{piddata} shows 13 TeV \pp\ PID spectrum data (points) as densities on \pt\ vs transverse rapidity \yt\ with $\pi$ mass assumed. Especially for $\Xi$s note that there is no other rescaling. Solid curves are full TCM parametrizations.  
Dashed curves are TCM soft components in the form $ z_{si }\bar \rho_s \hat S_0(y_t)$. 
The dotted curve in (d) is $\Omega$ hard-component model $z_{hi}(n_s) \bar \rho_h \hat H_{0i}(p_t)$ indicating that {\em almost all detected \pp\ $\Omega$s are jet fragments.}

A TCM including only an invariant soft component describing participant-nucleon dissociation and a slowly-varying hard component quantitatively related to jet production accurately describes PID spectra for heavy-hadron species from \pp\ and \ppb\ collisions. There is no need for a flow element or indication of a dense medium.

\section{heavy hadron $\bf \bar p_t$ trends} \label{mptsec}

This section reviews TCM analysis of PID ensemble-mean \mmpt\ data for multistrange hadrons as reported in Ref.~\cite{ppsss} based on PID spectra from Refs.~\cite{aliceppbpid,alippbss,didiersss} for 5 TeV \ppb\ collisions and from Refs.~\cite{alicepppid,alippss} for 13 TeV \pp\ collisions.   \mmpt\ trends for \pp\ and \ppb\ collision systems vs hard/soft ratio $x\nu$ are similar and relate simply to two basic fragmentation processes described by the TCM.  

This presentation is motivated in part by claims that \mmpt\ data from high-energy nuclear collisions serve as temperature measures related to thermal physics and equilibration: ``We predict that the mean transverse momentum of charged hadrons rises as a function of the charged-particle multiplicity in ultracentral nucleus-nucleus collisions. We explain that this phenomenon has a simple physical origin {\em and represents an {unambiguous prediction} of the hydrodynamic framework of heavy-ion collisions} [emphasis added]''~\cite{gardim}. Similarly, fluctuations of {\em event-wise} mean \pt\ or $\langle p_t \rangle$ are claimed to reflect hydrodynamic properties~\cite{parida}. Mean-\pt\ fluctuations have been studied in Refs.~\cite{ptscale,ptedep} where they are directly related to angular correlations arising from jet production. Various aspects of event-wise and ensemble \mmpt\ data are directly coordinated with jet production which  falsifies thermalization claims. This section considers \mmpt\ trends for heavy hadrons in small systems  relating to jet production.

\subsection{PID TCM for ensemble-mean $\bf \bar p_t$}

Given the PID spectrum TCM in Eq.~(\ref{pidtcmm}) the corresponding ensemble-mean {\em total} \pt\ for identified hadrons of species $i$, integrated over some angular acceptance $\Delta \eta$ that includes total integrated charge $n_{ch} = n_s + n_h$, is
\bea \label{ptintid}
\bar P_{ti}(n_{ch}) &=& \int_0^\infty dp_t' p'^{2}_t \bar \rho_{0i}(p_t',n_{ch}).
\\ \nonumber
&=&  n_{si}(n_s) \bar p_{tsi} +  n_{hi}(n_s) \bar p_{thi}(n_s).
\eea
Ensemble-mean $\bar p_{ti}$ is conventionally defined in terms of PID spectrum $\bar \rho_{0i}(p_t)$ in TCM form as 
\bea \label{oldmmpt}
\bar p_{ti}(n_{ch}) &=& \frac{\bar P_{ti}(n_{ch})}{n_{chi}} \approx \frac{\bar p_{tsi} + \tilde z_i x \nu \bar p_{thi}}{1 + \tilde z_i x \nu},
\eea
where factor $n_{si}$ has been canceled in the ratio and mean values $\bar p_{tsi}$ and $\bar p_{thi}$ are derived from TCM model functions required to describe spectrum data. As with a number of conventional data presentation formats this format conceals an underlying simplicity.
A simpler mean-value measure may be defined by
\bea \label{tcmmmpt}
\frac{\bar P_{ti}(n_{ch})}{n_{si}} \approx \bar p_{tsi} + \tilde z_i x \nu \bar p_{thi},
\eea
permitting straightforward inference of $\bar p_{thi}$ from $\bar p_{ti}$-vs-$x\nu$ data for each hadron species and collision system given factors $\tilde z_i$ and terms $\bar p_{tsi}$ as presented in Ref.~\cite{ppsss}.

\subsection{Ensemble-mean $\bf \bar p_t$ for multistrange hadrons} \label{strangemmpt}

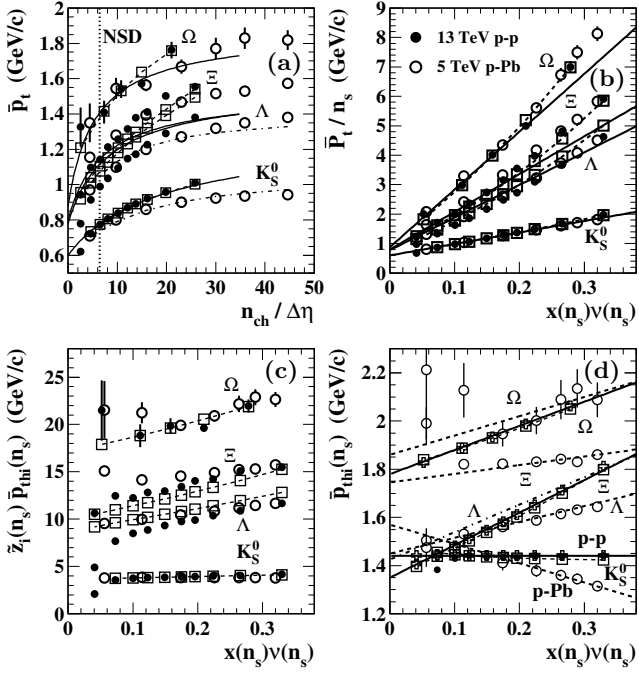
\begin{figure}[h]
	\includegraphics[width=3.3in]{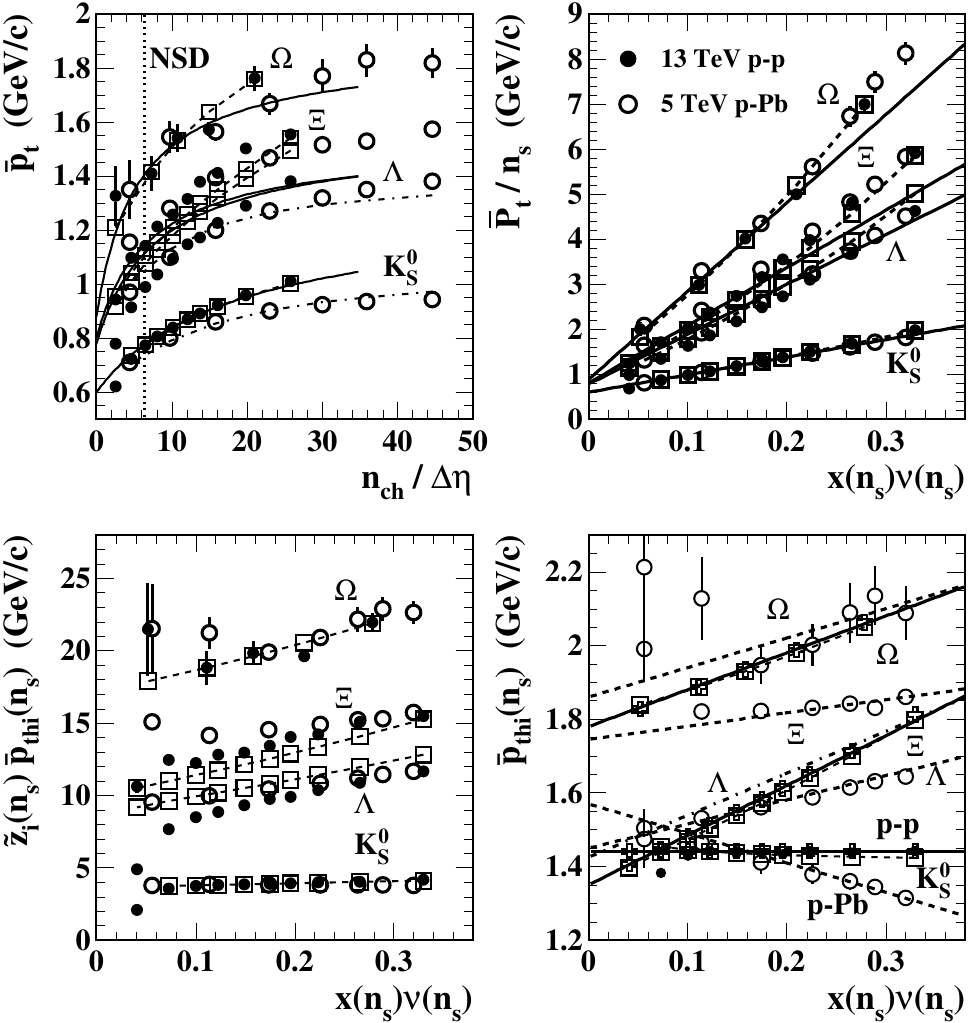}
	\put(-140,225) {\bf (a)}
	\put(-19,219) {\bf (b)}
	\put(-141,109) {\bf (c)}
	\put(-21,109) {\bf (d)}
	\\
	\caption{\label{mptss}
		(a) Ensemble-mean \mmpt\ data for 13 TeV \pp\ collisions from Fig.~5 of Ref.~\cite{alippss} (solid dots) and for 5 TeV \ppb\ collisions (open circles) from Ref.~\cite{aliceppbpid}. Open squares (connected by dashed curves) are \pp\ \mmpt\ values inferred from TCM curves in Fig.~\ref{piddata}. 
		Dash-dotted curves are \ppb\ \mmpt\ TCM trends from Refs.~\cite{ppbpid,pidpart2}. 
		(b) Data and curves from (a) corresponding to Eq.~(\ref{oldmmpt}) transformed to the TCM format of Eq.~(\ref{tcmmmpt}).
		(c) Quantity $\tilde z_i(n_s) \bar p_{thi}(n_s)$ derived from data and curves in (b) based on estimates of $\bar p_{tsi}$ and TCM hard/soft ratio $x(n_s)\nu(n_s)$.
		(d) Hard components $\bar p_{thi}(n_s)$ derived from data in (c) based on $\tilde z_i(n_s)$ trends described by Eq.~(\ref{zxi}). The open circles are \k0s\, $\Lambda$, $\Xi$ and $\Omega$ data from 5 TeV \ppb\ collisions. Open squares and dashed curves are derived from the \pp\ variable TCM. Open crosses are derived directly from corresponding TCM hard-component model functions $\hat H_0(y_t,n_s)$. 	Solid dots correspond to 13 TeV \pp\ \k0s\ data. The dash-dotted line represents the TCM for 13 TeV \pp\ $\Lambda$s.
	}  
\end{figure}

Figure~\ref{mptss} shows the \mmpt\ TCM applied to multistrange hadron data. 
Data for \k0s\, $\Lambda$, $\Xi$ and $\Omega$ from 13 TeV \pp\ collisions (solid dots in all panels) are as reported in Ref.~\cite{alippss}. Open circles from 5 TeV \ppb\ collisions are \k0s\ and $\Lambda$ data as reported in Ref.~\cite{aliceppbpid} and $\Xi$ and $\Omega$ data as reported in Ref.~\cite{didiersss}. Solid curves correspond to a \pp\ PID TCM with fixed hard components corresponding to event class 5 (except  class 3 for $\Omega$). Open squares (connected by dashed curves) are \mmpt\ values inferred from  {\em variable}-TCM PID spectra (solid curves in Fig.~\ref{piddata}).

Figure~\ref{mptss} (a) presents a conventional \mmpt\ format per Eq.~(\ref{oldmmpt}). Panel (b) presents the same data and curves transformed to the format of Eq.~(\ref{tcmmmpt}) by multiplication with factor $1 + \tilde z_i(n_s) x(n_s) \nu(n_s)$ ($\nu \rightarrow 1$ for \pp\ data). In the format of Eq.~(\ref{tcmmmpt}), panel (b), \pp\ and \ppb\ data are directly comparable, and the close correspondence between the two systems (note open circles vs solid dots) is evident whereas that is not the case in panel (a).

Panel (c) shows a further manipulation via Eq.~(\ref{tcmmmpt}) so that the product $\tilde z_i(n_s) \bar p_{thi}(n_s)$ is isolated given estimates of $\bar p_{tsi}$ from TCM model functions $\hat S_{0i}(y_t)$. 
For each hadron species the same transformation is applied to data and curves. The general increases vs $x(n_s) \nu(n_s)$ can be attributed to {\em shifts} of baryon hard components to higher \yt\ and {\em mass-dependent} increases of hard/soft ratios $\tilde z_i(n_s)$. The \pp\ \k0s\ data (solid dots) show good agreement with the TCM and seem to agree with \ppb\ data in that format. However, there appear to be significant deviations from the TCM for \pp\ $\Lambda$ and $\Xi$ data. 

Panel (d) isolates $\bar p_{thi}(n_s)$ via division of panel (c) by mass-dependent $\tilde z_i(n_s)$. For the TCM that is a consistency check. The open squares are derived from TCM spectra (solid curves in Fig.~\ref{piddata}) while the open crosses are derived directly from $\hat H_0(y_t)$ model functions. If TCM analysis were linear and self-consistent the two results should be essentially identical, and they are. The open circles are $\bar p_{thi}(n_s)$ values inferred by transforming 5 TeV \ppb\ $\bar p_{ti}$ data from panel (a) as described above, with statistical errors only. The solid dots correspond to 13 TeV \pp\ \k0s\ data from Ref.~\cite{alippss}. The dash-dotted line represents TCM results for $\Lambda$s from 13 TeV \pp\ collisions. This is a highly differential  format with precision $\approx 30$ MeV/c or 2\% of $\bar p_{thi}$ averages over event classes.

Much of the information carried by PID \mmpt\ data is represented in panel (c).
These slopes for \mmpt\ trends in panel (b) increase by factor 30 with hadron mass, $\approx 0.7$ GeV/c for $\pi$s vs $\approx 20$ GeV/c for $\Omega$s, by the combination of $\tilde z_i(n_s)$ (simply proportional to hadron mass per Fig.~\ref{tildezparamss}) and $\bar p_{thi}(n_s)$ (via spectrum hard components reflecting PID fragmentation functions~\cite{ppbpid}). Variation of $\tilde z_i(n_s)$ shows no sensitivity to baryon identity or strangeness content {\em per se}, and given panel (d) there is no clear systematic trend for $\bar p_{thi}(n_s)$ other than mesons vs baryons. 
The simplicity here demonstrates the importance of a TCM reference for precision tests of data and models. 

\subsection{Direct comparison: jet fragments vs mean $\bf \bar p_t$} \label{directcomp}

As demonstrated in Fig.~\ref{ratioss} (right) hadron species abundances are similar for \pp\ and \ppb\ collisions when compared vs hard/soft ratio $x(n_s)\nu(n_s)$. Soft components $\hat S_{0i}(y_t)$ vary simply with mass, independent of collision system. Since \mmpt\ soft components $\bar p_{tsi}$ are derived from $\hat S_{0i}(y_t)$ the same applies to them. Thus, relevant variations of $\bar p_t(n_{ch})$ arise only from evolution of {\em jet-related} hard components $\bar p_{thi}(y_t,n_s)$. Especially for higher-mass hadrons with limited \pt\ coverage exponent $q$ is not important; $q$ is fixed at 3.7 for kaons and 4.6 for baryons. Modes $\bar y_{ti}$ and widths $\sigma_{y_ti}$ are the critical parameters.

\begin{figure}[h]
	\includegraphics[width=3.3in]{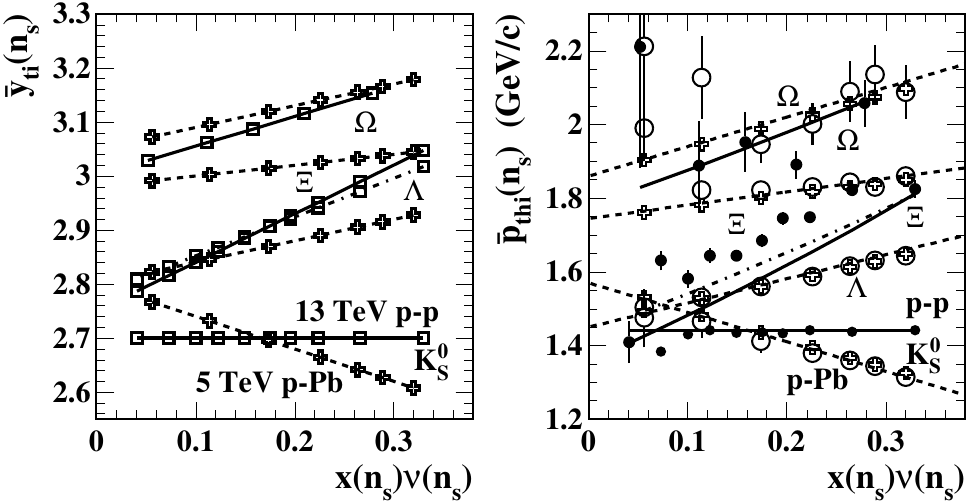}
	\caption{\label{ybarcomp}
		Left: Variable hard-component modes $\bar y_t(n_s)$ vs hard/soft ratio $x(n_s)\nu(n_s)$ for four hadron species and for 13 TeV \pp\ collisions (solid lines, open boxes) and 5 TeV \ppb\ collisions (dashed lines, open crosses). The trends are described in detail in Sec.~III of Ref.~\cite{ppsss}.
		Right: Corresponding TCM $\bar p_{thi}(n_s)$ trends obtained from TCM models $\hat H_{0i}(y_t)$ (lines, open crosses) with the same plot conventions. Also shown are $\bar p_{thi}(n_s)$ trends obtained from PID \mmpt\ data as described for Fig.~\ref{mptss} (d) for \ppb\ collisions (open circles) and \pp\ collisions (solid dots). The dash-dotted line corresponds to the 13 TeV \pp\ $\Lambda$ TCM.
		}  
\end{figure}

Figure~\ref{ybarcomp} shows a comparison between TCM hard-component  modes $\bar y_{ti}(n_s)$ (left) and \mmpt\ hard components $\bar p_{thi}(n_s)$ (right). Hard-component mode shifts at left are described in detail in Sec.~III of Ref.~\cite{ppsss}. The 13 TeV \pp\ $\Lambda$ mode trend is indicated by the dash-dotted line to distinguish from the \pp\ $\Xi$ trend (solid).

Figure~\ref{ybarcomp} (right) is Fig.~\ref{mptss} (d) with supplementary information. It shows $\bar p_{thi}(n_s)$ trends derived from  \mmpt\ data (solid dots and open circles for \pp\ and \ppb\ collisions respectively) corresponding to $\bar y_t(n_s)$ at left. 
Data are in this case plotted with  statistical uncertainties only. The dash-dotted curve represents \pp\ $\Lambda$s. Open crosses are derived directly from \ppb\ TCM hard components. 

Small $\bar y_{ti}(n_s)$ vs  $\bar p_{thi}(n_s)$ differences in relative position arise mainly from differences in TCM hard-component widths $\sigma_{y_ti}$ that also affect the resulting $\bar p_{thi}(n_s)$ trends.  $\bar p_{thi}(n_s)$ trends for  \k0s\ at right are higher relative to other species than $\bar y_{ti}(n_s)$ trends at left because $\sigma_{y_ti}$ for kaons is 0.58 (above the mode) compared to 0.50 for other species (except $\Xi$s as noted below). The \pp\ TCM trend for $\Xi$s $\bar p_{thi}(n_s)$ at right (solid line) is lower relative to other species compared to its $\bar y_{ti}(n_s)$ trend at left because {\em for \pp\ spectrum data only} the required $\Xi$ width $\sigma_{y_ti}$ is 0.46 rather than 0.50 (see below).

$\bar p_{thi}(n_s)$ values derived from \mmpt\ data generally agree with TCM trends within statistical uncertainties, with the exception of \pp\ $\Xi$s. ($\Lambda$ data for \pp\ are not plotted so as to minimize confusion among different symbols.) \pp\ $\Xi$ data in  the right panel (solid dots) lie systematically {\em above} the trend (solid line) derived from TCM spectra in Fig.~\ref{piddata}. 
As noted for $z_{0i}(n_s)$ trends, \pp\ $\Xi$ {\em spectra} are unique in requiring $z_{0i}^* = 0.00030$ rather than 0.00047 and $\sigma_{y_ti}=0.46$ rather than 0.50. Yet {\em integral measures} derived from the same $\Xi$ spectra such as $z_{0i}(n_s)$ values in Fig.~\ref{ratioss} (right)  conform to expectations from other species and to the TCM. 

Note that correspondence between the PID TCM and details of PID \mmpt\ data at the percent level and correspondence between left and right panels in Fig.~\ref{ybarcomp} rely on a TCM based on accurate determination of jet contributions to PID spectra and on accurate determination of \ppb\ centrality and hard/soft ratio $x(n_s)\nu(n_s)$~\cite{tomglauber}.

In this highly-differential array of jet-related spectrum parameters there is no indication of trend discontinuities or phase transitions. Parameter trends are consistent to a few percent over the full range of event classes for \pp\ and \ppb\ collisions and there is no evidence for flows.

\section{Strangeness enhancement} \label{enhance}

Strangeness enhancement (SE) relates to {\em fractional abundances} $z_{0i}(n_s)$ in the expression $\bar \rho_{0i} = z_{0i}(n_s) \bar \rho_0$ -- i.e.\ what fraction of total particle density $\bar \rho_0$ is represented by density $\bar \rho_{0i}$ for hadron species $i$. In the present context, is $z_{0i}(n_s)$ greater than expected from statistical-model predictions~\cite{statmodel} for some experimental conditions? This section summarizes Sec.~IV of Ref.~\cite{ppsss}.

SE is conventionally interpreted as evidence for QGP formation assuming that strange quarks  become more abundant in a high-temperature QCD medium; abundances of hadrons including strange quarks are then expected  to be enhanced~\cite{rafelski}. Supporting arguments are  based on a near-monolithic particle-production medium described by thermodynamics and hydrodynamics. Jets (assumed confined to high \pt) should play a negligible role. However, detailed analysis of spectrum structure and related strangeness properties reveals strong jet effects that may be mistaken for QGP-related effects.

In the plots below the preferred independent variable is $x\nu$, the hard/soft ratio that measures the fraction of total produced particles associated with jets. That preference follows from material presented in Secs.~\ref{structure} and \ref{mptsec} that demonstrate the dominant importance of jet production.
The conventional preference $\bar \rho_0 = n_{ch} / \Delta \eta = \bar \rho_s + \bar \rho_h$ mixes jet and nonjet hadron production mechanisms.

\subsection{Fractional abundances $\bf z_{si}(n_s)$ and $\bf z_{hi}(n_s)$}

For the PID TCM described in App.~\ref{tcmpid} soft and hard fractional abundances $z_{si}(n_{ch})$ and $z_{hi}(n_{ch})$ play a critical role.
In Ref.~\cite{pidpart1} $z_{si}(n_{ch})$ and $z_{hi}(n_{ch})$ are measured for seven event classes of 5 TeV \ppb\ collisions. $z_{si}(n_{ch})$ is expressed by Eq.~(\ref{zxi}) given Eq.~(\ref{pidtcmm}). Based on measured values for $z_{xi}$ the ratios $\tilde z_i(n_{ch}) \equiv z_{hi}(n_{ch}) / z_{si}(n_{ch})$ are inferred as in Fig.~\ref{tildezparamss} (left). The limiting case of Eq.~(\ref{zxi}) for $x\nu \rightarrow 0$ is $z_{si}(0) \rightarrow z_{0i}(0)$. More generally,
\bea
z_{0i}(n_{ch}) &=& \frac{z_{si}(n_{ch})+ x\nu z_{hi}(n_{ch})}{1 + x\nu}.
\eea
Measured values of $z_{xi}$ for $\pi$s, kaons, protons and $\Lambda$s indicate that ratio $z_{0i}(n_{ch}) / z_{0i}(0)$ remains {\em within a few percent of unity} for all \nch. That is, $z_{0i}(n_{ch})$ inferred from $z_{xi}(n_{ch})$ measurements maintains a fixed value vs $x\nu$ within data uncertainties as demonstrated by $z_{0i}(n_{ch}) / z_{0i}(0)$ in Fig.~10 (right) of Ref.~\cite{pidpart1}.

That procedure compares an ideal TCM model of nonPID data spectra described within data uncertainties to TCM models of PID data spectra, also within uncertainties. In contrast to isolated data spectra, TCM models are strongly constrained as to $\hat S_{0i}$ and $\hat H_{0i}$ model structures and thus span a large \pt\ acceptance. There is no need for interpolations and extrapolations beyond limited acceptance intervals. That is especially important for $\Xi$ and $\Omega$ data with their limited acceptances.

\subsection{Apparent SE evidence from spectrum integrals}

\begin{figure}[h]
	\includegraphics[width=3.3in]{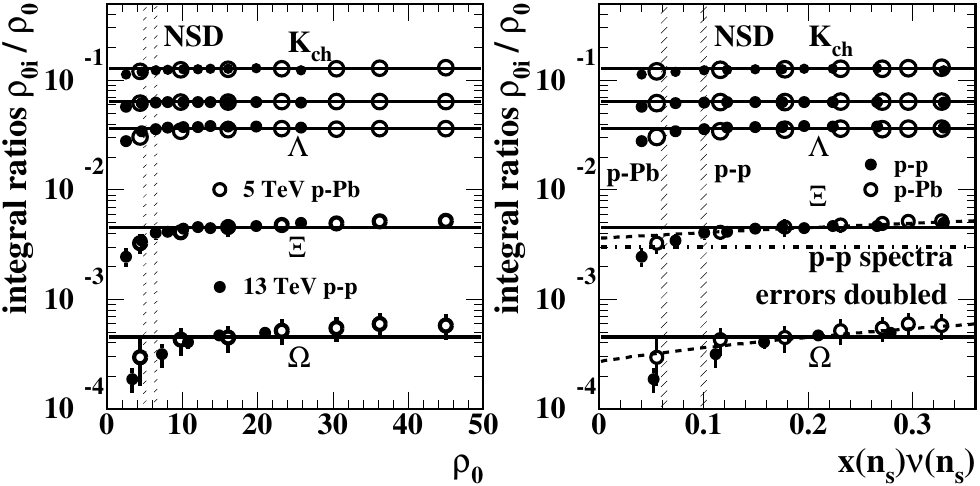}
	\caption{\label{ratioss}
		Left: Integral ratios $z_{0i}(n_s) \equiv \bar \rho_{0i} / \bar \rho_0$ vs $\bar \rho_0$ based on 13 TeV \pp\ data in Fig.~8 of Ref.~\cite{alippss} (solid dots) and 5 TeV \ppb\ data in Table~4 of Ref.~\cite{alippbss} (open circles) which then represent variable quantity  $z_{0i}(n_s)$ in the TCM context. $\Xi$ and $\Omega$ errors have been doubled to make them visible. The solid lines are fixed values $z_{0i}^*$ as reported in the present study. The hadron species, from the top, are charged kaons, \k0s, $\Lambda$s, $\Xi$s and $\Omega$s.
		Right: Data in the left panel plotted vs TCM hard/soft ratio $x(n_s) \nu(n_s)$ demonstrating the near equivalence of \pp\ and \ppb\ collision systems for quantity $z_{0i}(n_s)$ plotted vs that ratio. The dash-dotted line corresponds to \pp\ $\Xi$ spectra in Fig.~\ref{piddata} (c). 
	}  alippss4a
\end{figure}

Figure~\ref{ratioss} (left) compares \pp\ data (solid dots) and \ppb\ data (open circles) for integrated yields $\bar \rho_{0i}$ in the form of ratios $ z_{0i}(n_s) \approx \bar \rho_{0i} / \bar \rho_{0}$. Data sources are reported in Ref.~\cite{ppsss}.
The error bars show systematic errors uncorrelated across event classes. 
The hatched bands represent NSD values $\bar \rho_{0\text{NSD}} \approx 5$ for 5 TeV \ppb\ collisions and $\approx 6.5$ for 13 TeV \pp\ collisions. 

Figure~\ref{ratioss} (right) shows results in the left panel replotted vs hard/soft ratio $x(n_s) \nu(n_s)$ (with $\nu = 1$ for \pp). In that plot format the near equivalence of \ppb\ and \pp\ data is apparent.  A similar equivalence between \pp\ and \ppb\ systems is revealed for ensemble-mean \mmpt\ data in Sec.~\ref{mptsec}. In either case the equivalence reflects the dominant role of jets in the measured phenomena. $\bar \rho_0$ is composite, representing two hadron production mechanisms additively, whereas $x\nu$ is a hard/soft ratio ensuring informative comparisons among different A-B systems.
$\Omega$ and $\Xi$ errors are doubled for visibility. 

Significant variation of parameters $z_{0i}(n_s)$ for $\Xi$ and $\Omega$ baryons is evident in Fig.~\ref{ratioss}. In order to incorporate that feature into a PID TCM, yield data should be examined more differentially.
Variable $z_{0i}(n_s)$ is  expressed in terms of hard/soft ratio $x \nu$ as 
\bea \label{z0trend}
z_{0i}(n_s) \approx z_{0i}^* [1 + \delta z_{0i}^*(x\nu - 0.20)].
\eea
Value 0.20 is a consequence of choosing $z_{0i}^*$ values corresponding to spectra for central event class 5 (for \pp) or 4 (for \ppb). The crossover in Fig.~\ref{ratioss} (right) then occurs near $x\nu = 0.2$.  
$z_{0i}^*$ and $\delta z_{0i}^*$ values are given in Ref.~\cite{ppsss}.

\begin{figure}[h]
	\includegraphics[width=3.3in]{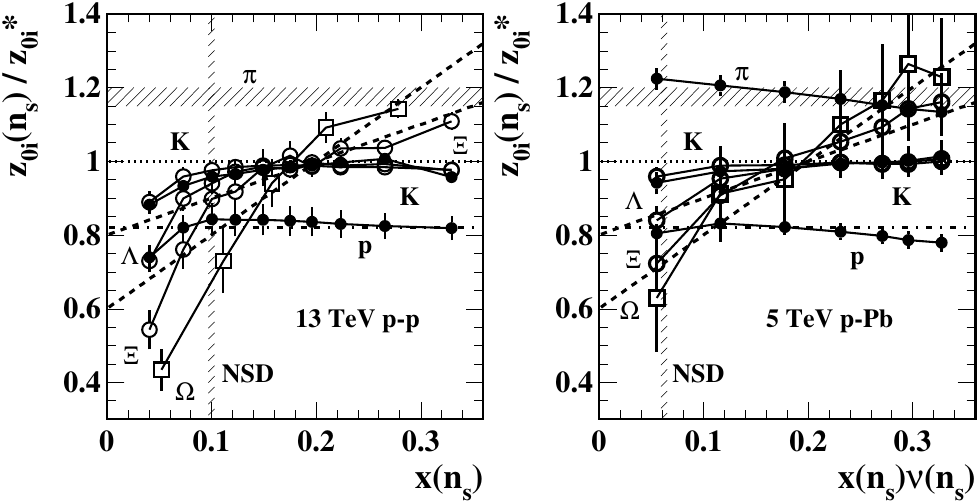}
	\caption{\label{ratiosx}
		Rescaled hadron fraction coefficients in the form $z_{0i}(n_s) / z_{0i}^*$ for 13 TeV \pp\ collisions (left) and 5 TeV \ppb\ collisions (right) and for charged hadrons (solid dots) and neutral hadrons (open circles, except open boxes for $\Omega$s). The vertical hatched bands denote hard/soft ratio values corresponding to NSD collisions. The horizontal hatched bands estimate the biased $z_{0i}(n_s) / z_{0i}^* \approx 1.17$ value noted for \pp\ $\pi$s. The dash-dotted lines are the average value of the proton inefficiency correction function of  Eq.~(10) in Ref.~\cite{pppid}.
	} 
\end{figure}

Figure~\ref{ratiosx} presents a more differential analysis of coefficients $z_{0i}(n_s)$ showing ratios $z_{0i}(n_s) / z_{0i}^*$ vs hard/soft ratio $x(n_s)$ for \pp\ collisions (left) and vs $x(n_s)\nu(n_s)$ for \ppb\ collisions (right) and for charged hadrons (solid dots) and neutral hadrons (open circles, or open squares for $\Omega$s) where curves guide the eye. In the right panel it is notable that the {\em published} yields for $\pi$s and protons relative to statistical-model expectations differ systematically from reference value 1 to an extent consistent with bias arising from $dE/dx$ PID crosstalk as discussed in Refs.~\cite{pidpart1,pppid}. The dash-dotted lines represent the {\em average} value 0.82 of proton inefficiency function $\epsilon_p(p$-$p)$ defined by Eq.~(10) of Ref.~\cite{pppid}. The hatched band for $\pi$s including 1.17 complements the proton trend.

All hadron species decrease substantially relative to reference value 1 for \nch\ {\em below} NSD values (e.g.\ $\bar \rho_0 \approx 6.5 \Rightarrow x \approx 0.1$ for \pp\ collisions, vertical hatched band). No attempt is made within the TCM to describe such {\em local} low-\nch\ trends. The {\em dashed lines} in the two panels describe a more-gradual variation approximately linear vs hard/soft ratio $x$ or $x\nu$. Slopes of the dashed lines are then incorporated (as $\delta z_{0i}^*$) into Eq.~\ref{z0trend}. The dashed lines for $\Xi$ and $\Omega$ pass through 1.0 for $x\nu \approx 0.2$. 

Discounting suppression below NSD charge densities common to all hadron species, only $\Xi$s and $\Omega$s manifest significant increases linear on hard/soft ratio $x\nu$ for two collision systems. Whether those trends relate to real SE effects is not clear from these data.

\subsection{Apparent SE trends vs jet production} \label{highpt}

This subsection relates specifically to Fig.~6 of Ref.~\cite{alippss} where high-\pt\ integrals of spectra for \k0s, $\Lambda$s, $\Xi$s and $\Omega$s are plotted vs rescaled event multiplicity. Those data trends conceal important information.

\begin{figure}[h]
	\includegraphics[width=3.3in]{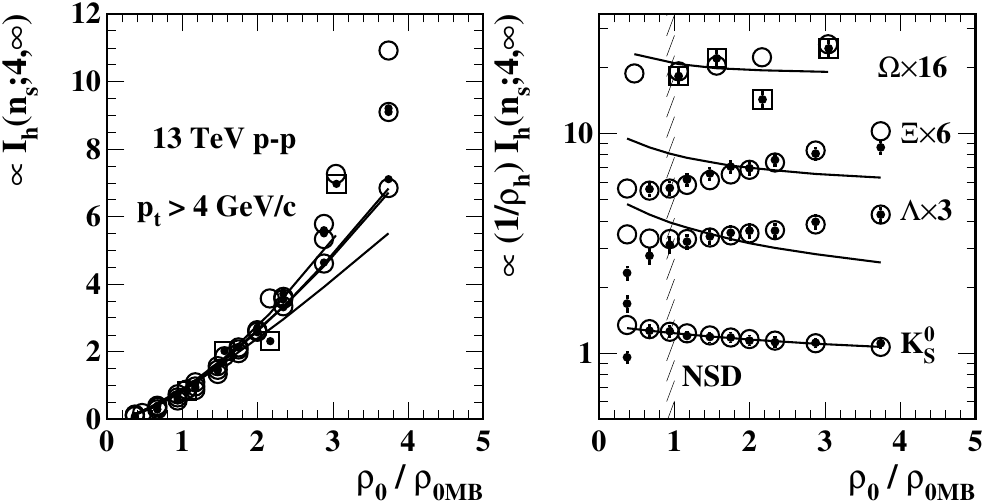}
	\put(-140,55) {\bf (a)}
	\put(-21,55) {\bf (b)}\\
	\includegraphics[width=1.65in]{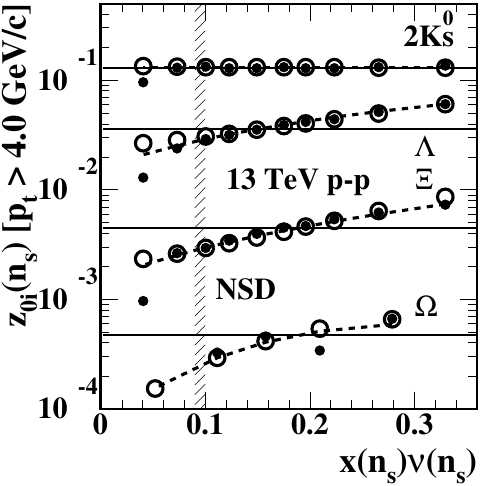}
	\includegraphics[width=1.62in]{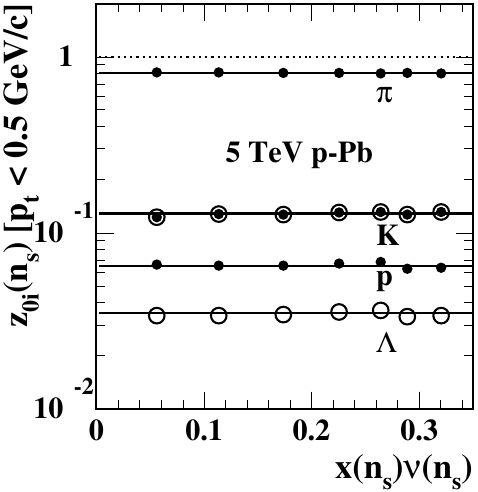}
	\put(-140,54) {\bf (c)}
	\put(-20,82) {\bf (d)}
	\caption{\label{highptx}
		(a) High-\pt\ PID spectrum integrals ($p_t > 4$ GeV/c) from 13 TeV \pp\ collisions (solid dots) for four hadron species as presented in Fig.~6 (top row) of Ref.~\cite{alippss}. Open squares identify $\Omega$ data. Open circles and curves are TCM results.
		(b) Data and curves in the left panel rescaled by $\bar \rho_h = x(n_s) \bar \rho_s$. Solid dots are data from Ref.~\cite{alippss}. Open circles are derived from TCM spectra (curves) in Fig.~\ref{piddata}. Solid curves are TCM trends with no hard-component mode shift on \yt.
		(c) Contents $I_h / \bar \rho_h$ of panel (b) transformed according to Eq.~(\ref{panelc}).
		(d) A similar procedure applied to spectrum integrals over \pt\ interval [0,0.5] GeV/c showing no significant $z_{0i}$ variation.
	} 
\end{figure}

Figure~\ref{highptx} (a) shows PID yield data (solid dots) for 13 TeV \pp\ collisions from Ref.~\cite{alippss} corresponding to its Fig.~6 (upper left). PID spectra in Fig.~\ref{piddata} above (solid dots) are integrated over high-\pt\ intervals ($> 4$ GeV/c, except $>3.8$ GeV/c for $\Omega$s). Integrated yields are then rescaled by the same integral applied to a minimum-bias spectrum for the corresponding hadron species.  A similar rescaling is applied to total charge density $\bar \rho_0$ on the $x$ axis, where $\bar \rho_{0\text{MB}} \approx 6.9$. The open squares highlight the $\Omega$ solid points. Note in Fig.~\ref{piddata}  that there are only two $\Omega$ data points above 3.8 GeV/c for four event classes and only one for the lowest \nch\ (explaining a missing point in Fig.~\ref{highptx} (b)). The open circles result from the same procedure applied to TCM spectra (solid curves) in Fig.~\ref{piddata}. The solid curves in (a) are explained below. Given Eq.~(\ref{pidtcmm}) the integrals have the form
\bea \label{ratint}
I_h(n_s;4,\infty)&\equiv& z_{hi}(n_s) \bar \rho_{h} \int_\text{4 GeV/c}^\infty p_t dp_t \hat H_{0i}(p_t,n_s),~~~~
\eea
where $\bar \rho_{h} \approx \alpha \bar \rho_s^2 \equiv x(n_s) \bar \rho_s$ and $\bar \rho_s$ is obtained from measured charge density $\bar \rho_0$ as the root of $\bar \rho_0 \approx \bar \rho_s + \alpha \bar \rho_s^2$, with $\alpha \approx 0.017$ for 13 TeV \pp\ collisions~\cite{tomnewppspec}. Soft components $\hat S_{0i}(p_t)$ can be neglected above 4 GeV/c. Clearly, panel (a) does not provide visual access to useful data trends. 

The data increase in panel (a) relative to linear $\bar \rho_0$ is effectively dominated in Eq.~(\ref{ratint}) by factor $\bar \rho_h \propto \bar \rho_s^2 \sim \bar \rho_0^2$, a quadratic relation that is a {\em signature feature of jet production} in \pp\ collisions~\cite{jetspec2}. The \pt\ lower bound on the integral in Eq.~(\ref{ratint}) is well above hard-component modes (also excluding data soft components), and effectively represents integrals of the high-\pt\ tails of jet fragment distributions particularly sensitive to any mode or width variations of the peaked hard component.

Figure~\ref{highptx} (b) shows an alternative plotting format based on the TCM. The error bars represent uncorrelated systematic uncertainties. Given the structure of Eq.~(\ref{ratint}), data and curves in panel (a) may be divided by common density $\bar \rho_h$ to obtain detailed information. If data hard components $\hat H_{0i}(p_t)$ were independent of event class $(n_s)$ the integrals in  Eq.~(\ref{ratint}) should all be the same.  Variation vs $\bar \rho_0$ would then be determined solely by PID coefficients $z_{hi} = \tilde z_i z_{si}$ via Eq.~(\ref{zxi}) (solid curves in panel (b)). But as reported in Refs.~\cite{pidpart2,pppid} spectrum hard components $\hat H_{0i}(p_t,n_s)$ vary systematically with event class, meson functions shifting to {\em lower} \yt\ ( \ppb\ collisions only) while baryon functions shift to higher \yt\ with corresponding substantial variation of  integrals in Eq.~(\ref{ratint}). Trends for high-\pt\ integrals in  (b) should then correspond to trends for hard-component modes $\bar y_{ti}$ in Fig.~\ref{ybarcomp} (left) above.

Rescaling of TCM elements for Fig.~\ref{highptx} (b) is performed as follows. For each hadron species the TCM integrals (open circles, derived from solid curves in Fig.~\ref{piddata}) are rescaled to agree {\em on average} with  ALICE data (solid dots) as they appear in panel (b).  TCM hard-component modes $\bar y_t$ are then temporarily fixed at event class 5 (or 3 for $\Omega$s), making the integral factors in Eq.~(\ref{ratint}) invariant, and solid curves ($\propto z_{hi}(n_s)$) for each hadron species are rescaled to coincide with resulting TCM integrals (open circles). {\em Variable} TCM hard components are then restored. TCM points and curves are also back-transformed appropriately to appear in panel (a). 

Figure~\ref{highptx}  (c) shows a further simplification. The solid curves in panel (b) $\propto z_{hi}(n_s)$ (including factor $z_{0i}(n_s)$) serve as a reference for the integrated-spectrum trends (solid and open points). An estimator {\em proportional to}  $z_{0i}(n_s)$ can then be obtained via Eq.~(\ref{zxi}) from 
\bea \label{panelc}
z_{0i}(n_s) \sim \frac{1 + \tilde z_i(n_s) x \nu}{\tilde z_i(n_s) (1 + x\nu)} I_h(n_s;4,\infty) / \bar \rho_h
\eea
in which horizontal lines would emerge if the integrals in Eq.~(\ref{ratint}) are independent of event class. Panel (c) reveals that integration of the high-\pt\ portions of spectra {\em with intent to evaluate}  $z_{0i}(n_s)$ produces a misleading result because of evolution of spectrum hard components (mode shifts on \yt) with changing event conditions. Refer to Fig.~\ref{ybarcomp} (left). Since data (solid points) in panel (a) were rescaled with inaccessible factors~\cite{alippss} results in panels (c) are rescaled to agree {\em on average} with $z^*_{0i}$ values in Fig.~\ref{ratioss}.

Figure~\ref{highptx}  (d) shows the same procedure applied to low-\pt\ [0,0.5] GeV/c portions of \ppb\ spectra. Inferred $z_{0i}(n_s)$ trends are then independent of event class consistent with observed stability of data soft components. The results in Figure~\ref{highptx}  [(c), (d)] may be compared with those in Fig.~\ref{ratioss} (right). In the latter case significant variation is observed only for $\Xi$s and $\Omega$s whereas Fig.~\ref{highptx} (c) shows {\em comparable} strong variation for $\Lambda$s, $\Xi$s and $\Omega$s. The hard/soft ratio for hadron species is directly mass dependent ($\tilde z_i \propto m_i$), and Fig.~\ref{ratioss} represents integrals over the full \pt\ acceptance including hard {\em and} soft components. Thus, relative to $\Omega$s the large variations in panel (c) are strongly reduced for $\Xi$s and negligible for $\Lambda$s in Fig.~\ref{ratioss}.

\subsection{Conclusions re strangeness enhancement}

Determination of total fractional abundances $z_{0i}(n_s)$ depends on accurate spectrum integrals, but spectrum integrals for heavier hadrons are strongly challenged by \pt\ acceptance limitations and resulting need for spectrum extrapolations to provide integral estimates. The problem is exacerbated by the hard/soft abundance ratio $\tilde z_{i} = z_{hi}/z_{si} \propto$  hadron mass $m_i$. Examples may be seen for $\Xi$s and $\Omega$s in Figs.~\ref{ppbpiddata} and \ref{piddata}. For $\Xi$s, soft components are not fully covered by data for more-central spectra, but for $\Omega$s soft components are not well defined for any centrality. In effect, $\Omega$ spectra are determined solely by jet fragments. In that case extrapolations to lower \pt\ may be dominated by {\em hard-component shapes above the mode}. In addition there is the issue illustrated by Fig.~\ref{highptx}, sensitivity to shifts in the hard-component mode to higher \pt\ with increasing event \nch\ {\em for baryons}. There is thus considerable risk of substantially overestimating fractional abundances for heavier hadrons {\em especially for more-central collisions.}

\section{Two-particle correlations} \label{2dcorr}

Two-particle correlations on angles $(\eta,\phi)$~\cite{axialci,axialcd} and transverse mass $(m_{t1},m_{t2})$~\cite{mtmt} or transverse rapidity $(y_{ti},y_{t2})$~\cite{porter2,porter3} have been studied extensively since first RHIC operation at 130 GeV. Distinctions among jet-related correlations and other  mechanisms and dramatic dependence on like-sign vs unlike-sign charge combinations have been explored. Below is a summary of results pertaining  to jet- or color-dipole-related and color-quadrupole-related correlation analysis. This material has particular relevance to claims cited in Sec.~\ref{promise} relating to perceived appearance of ``collectivity'' and {\em absence of significant jet structure} for $p_t < 5$ GeV/c.

\subsection{2D angular correlations} \label{angcorrr}

\begin{figure}[h]
\includegraphics[width=1.5in]{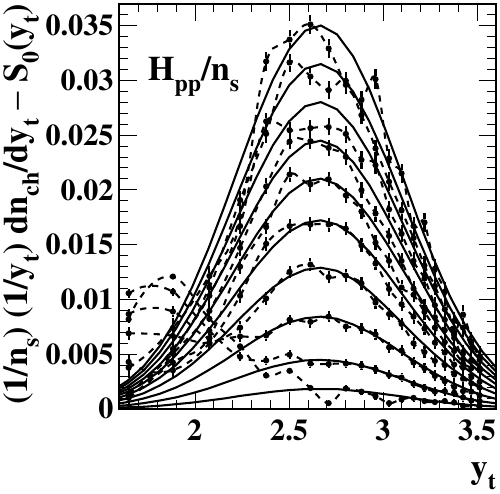}
\put(-21,94) {\bf (a)}
\includegraphics[width=1.65in]{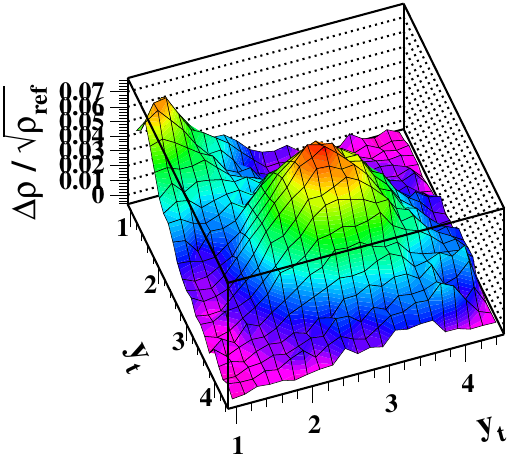}
\put(-15,94) {\bf (b)}\\
\includegraphics[width=1.65in]{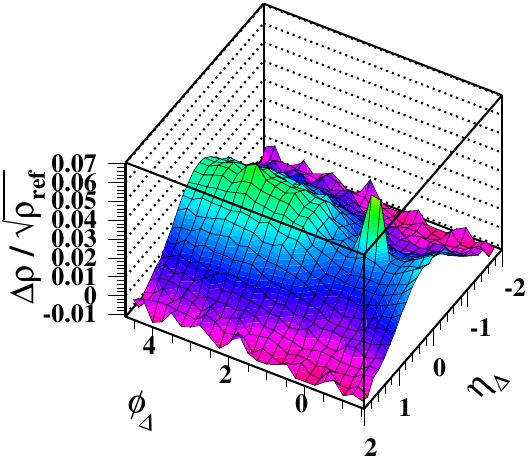}	
\put(-15,92) {\bf (c)}
\includegraphics[width=1.65in]{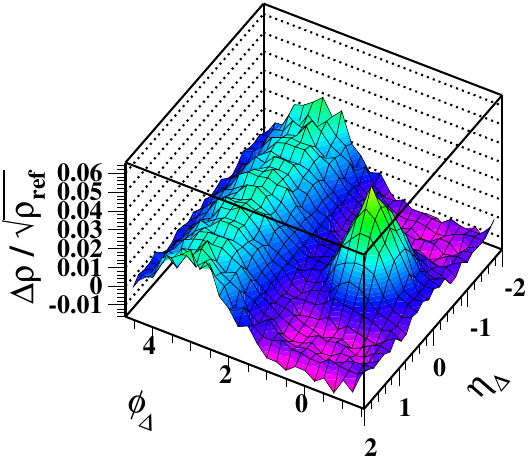}
\put(-15,92) {\bf (d)}
\caption{\label{ppcorrx}				
(a) Spectrum hard components from 200 GeV \pp\ collisions as in Ref.~\cite{ppprd}.
(b) Two-particle correlations on $(y_{t1},y_{t2})$ with soft and hard components.
Angular correlations on $(\eta_\Delta,\phi_\Delta)$ for (c) soft component and (d) hard component matching the two peaks in panel (b). ``Hard-component'' structure in panel (d) is limited to hadron pairs each with $p_t \approx 0.6$ GeV/c still demonstrating well-defined jet structure.
	} 
\end{figure}

Figure~\ref{ppcorrx} (a) shows spectrum hard components as in Fig.~\ref{piddataa} for ten multiplicity classes of 200 GeV \pp\ collisions plotted on transverse rapidity \yt. Figure~\ref{ppcorrx} (b) shows corresponding 2D correlations on space $(y_{t1},y_{t2})$ with distinct soft-component (low \yt) and hard-component (high \yt, jets) peaks. The latter as marginal projection matches the SP spectrum hard components in Fig.~\ref{ppcorrx} (a).

Figure~\ref{ppcorrx} (c) shows angular correlations corresponding to the soft-component peak in panel (b). Figure~\ref{ppcorrx} (d) shows angular correlations corresponding to the hard-component peak in (b) but with the added constraint that hadrons are selected from a small region on $(y_{t1},y_{t2})$ near (2,2) corresponding to $p_t \approx 0.6$ GeV/c, demonstrating that clear jet-like correlations persist at {\em very low} \pt.

\begin{figure}[h]
\includegraphics[width=1.65in]{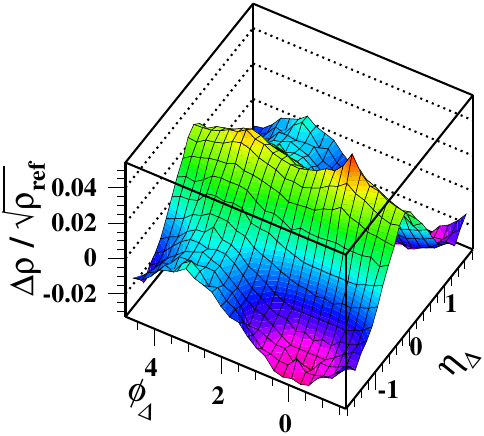}
\put(-27,97) {\bf (a)}
\includegraphics[width=1.65in]{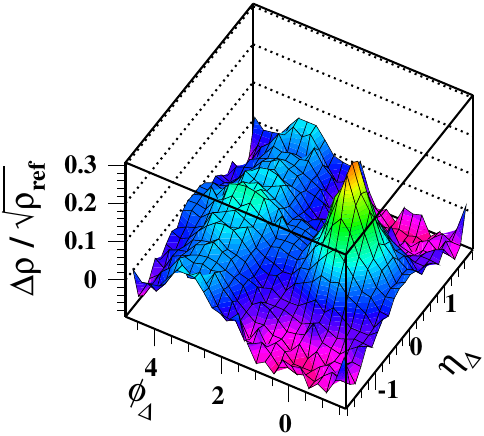}	
\put(-27,97) {\bf (b)}\\
\includegraphics[width=1.65in]{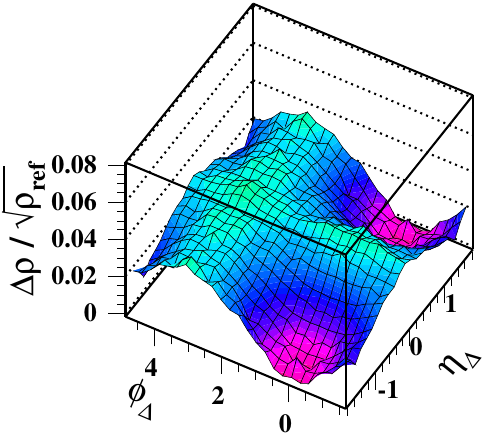}
\put(-27,97) {\bf (c)}
\includegraphics[width=1.65in]{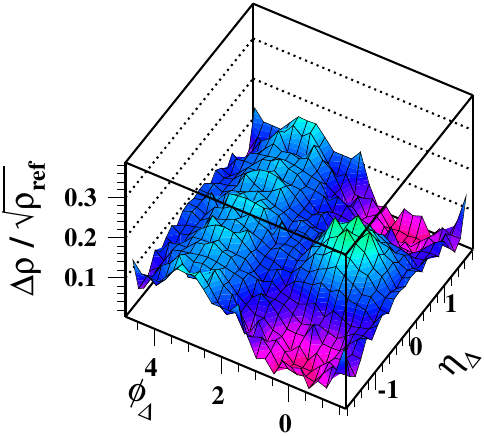}
\put(-27,97) {\bf (d)}
\caption{\label{ppcorr}				
Upper: 2D angular correlations from 200 GeV \pp\ collisions~\cite{ppquad} for low \nch\ (a) and high \nch\ (b).
Lower: Histograms as above but with fit-model elements subtracted to leave only dijet and nonjet quadrupole contributions (c,d).
	}  
\end{figure}

Figure~\ref{ppcorr} shows 2D angular correlations from 200 GeV \pp\ collisions~\cite{ppquad} as pair densities on angle differences $\eta_{\Delta}$ and $\phi_{\Delta}$ for \pp\ event classes 1 (a) and 6 (b) (of 7, see Fig.~\ref{ppquadjet}, left). The measured quantity is {\em per particle} correlations denoted by $\Delta \rho/\sqrt{\rho_{ref}}$ where $\sqrt{\rho_{ref}} \approx \bar \rho_0$ and $\Delta \rho$ represents numbers of correlated pairs (covariances). A six-element fit model applied to the measured histograms achieves data descriptions within point-point data uncertainties~\cite{ppquad}. Model amplitudes for this measure are denoted by $A_\text X$ as in $A_\text{2D}$ for  same-side 2D jet peak, $A_\text D$ for away-side dipole and $A_\text{Q}$ for nonjet quadrupole.

Figure~\ref{ppcorr} (c,d) shows data from (a,b) with model elements describing projectile-nucleon fragmentation (1D Gaussian on $\eta_\Delta$), Bose-Einstein correlations (narrow 2D exponential peak at the origin) and constant offset subtracted. What remain are contributions from minimum-bias jet production (same-side 2D peak, away-side dipole $\cos(\phi_{\Delta}-\pi)$) and a nonjet quadrupole element represented by $\cos(2\phi_\Delta)$. Those elements are described below.

\subsection{Color quadrupole angular correlations} \label{quadcorr}

\begin{figure}[h]

\includegraphics[width=1.65in]{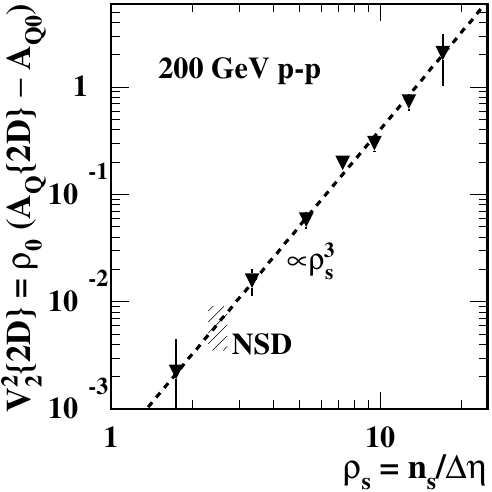}	\includegraphics[width=1.65in]{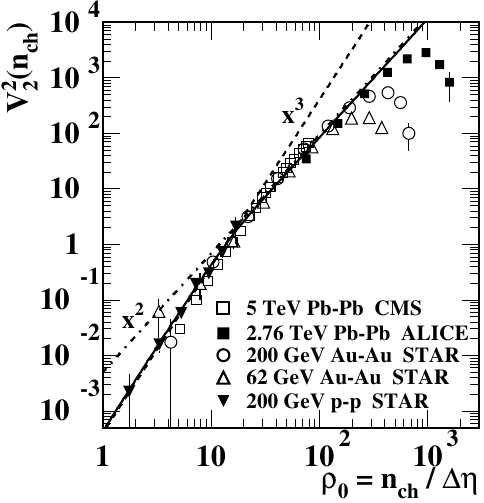}
	\caption{\label{ppquadjet}
	Left: Total correlated pairs for 200 GeV \pp\ azimuth quadrupole $V_2^2$ (solid triangles)
from Ref.~\cite{ppquad} vs trend $\propto \bar \rho_s^3$ (dashed).
	Right:  NJ quadrupole correlations from 62 and 200 GeV \auau\ collisions (open circles and triangles) from Ref.~\cite{anomalous} and 2.76 TeV \pbpb\ collisions (solid squares) from Ref.~\cite{alicev2b} as  $V^2_2$ vs charge density $\bar \rho_0$. {Also included are 5 TeV \pbpb\ $v_2(p_t)$ data (open squares) from Ref.~\cite{cmsabv2}.}  200 GeV \pp\ data from the left panel (solid triangles) are superimposed with their cubic trend (dashed). The solid curve is Eq.~(\ref{v22b}) rescaled to best match \pp\ and \auau\ data at lower \nch. The dash-dotted line $\propto \bar \rho_0^2$ is discussed in the text.	
}  
\end{figure}

Figure~\ref{ppquadjet} (left) shows quadrupole amplitudes (pair numbers) $V_2^2\{\text{2D}\} \equiv \bar \rho_0 A_\text{Q}\{2\text D\}$ (solid triangles) vs $\bar \rho_s$, where \{2D\} denotes data inferred from 2D model fits vs other $v_2$ ``methods''~\cite{ppquad}. $\bar \rho_s$ is the soft-component density in $\bar \rho_0 = \bar \rho_s + \bar \rho_h$ which may be interpreted as proportional to the density of low-$x$ gluons.  \pp\ data reveal that the azimuth quadrupole amplitude (as pairs) increases as  $V_2^2 \propto \bar \rho_s^3$, suggesting that color-quadrupole production arises from {\em three-gluon} interactions.

Figure~\ref{ppquadjet} (right) shows quadrupole data for \aa\ systems plotted as $V_2^2 \equiv \bar \rho_0^2 v_2^2$ with $\bar \rho_0$ and $v_2$ as published with no rescaling.  ALICE 2.76 TeV \pbpb\ $v_2(n_{ch})$ data are from Ref.~\cite{alicev2b}, ALICE $\bar \rho_0(n_{ch})$ data are from Ref.~\cite{alicerho0}, STAR \auau\  $\bar \rho_0(n_{ch})$ data are from Tables~III and IV of Ref.~\cite{anomalous} and STAR \auau\ $v_2(n_{ch})$ and $A_\text{Q}(n_{ch})$ data are from Ref.~\cite{v2ptb}. 
CMS 5 TeV \pbpb\ $v_2(n_{ch})$ data are from Ref.~\cite{cmsabv2}.
200 GeV \pp\ data and cubic trend (dashed line) are copied from the left panel.
Quadrupole correlation amplitude $V_2^2(\bar \rho_0)$ (as correlated pairs) increases by {\em almost six orders of magnitude} consistently across five collision systems. Again, there is no rescaling of data to achieve that simple correspondence.

{Given the \pp\ quadrupole data trend in Fig.~\ref{ppquadjet} (left) a conjecture $V_2^2(\bar \rho_0) \propto \bar \rho_s^3 \rightarrow \bar \rho_s \bar \rho_h$ may be tested~\cite{tomnewquad}. By analogy with that \pp\ relation the quadrupole trend for \aa\ data with \aa\ geometry parameters may be approximated by}
\bea \label{v22b}
 V_2^2(n_{ch}) &\equiv& \bar \rho_0^2 v_2^2(n_{ch}) \propto \bar \rho_s(n_{ch}) \times \bar \rho_h(n_{ch})
\\ \nonumber
&\propto& (N_{part}/2) \bar \rho_{sNN} \times N_{bin}  \bar \rho_{sNN}^2,
\eea
where \aa\ geometry parameters are derived from a TCM analysis of 2.76 TeV \pbpb\  data reported in Ref.~\cite{tompbpb} and summarized in App.~\ref{geometry}. 
Equation~(\ref{v22b}) is represented by the solid curve at right rescaled to best match \pp\ data at lower \nch. 
That {\em predicted} trend follows the cubic dashed line  $\propto \bar \rho_0^3$ at smaller $\bar \rho_0$ determined by \pp\ data wherein $N_{part}/2 \approx N_{bin} \approx 1$ because of {\em exclusivity}~\cite{tomexclude}. The trend is then determined by product $ \bar \rho_{sNN} \bar \rho_{hNN} \rightarrow \bar \rho_s^3$. At larger $\bar \rho_0$ the $ \bar \rho_{xNN}$ become approximately constant, and the trend is determined by product $N_{part} N_{bin} \propto \bar \rho_0^2$ (dash-dotted line) as determined in App.~\ref{geometry}, Fig.~\ref{story}.
The comparison confirms that $V_2^2$ data for A-B systems may be described in general by expression $\bar \rho_s \times \bar \rho_h$ (three-gluon interaction) except for high charge densities $\bar \rho_0$ where large {\em event-wise} numbers of {\em random} (on azimuth) quadrupole orientations may result in lower mean values.

The quadrupole trend in Fig.~\ref{ppquadjet} (left) indicates that quadrupole amplitudes measured as number of correlated pairs (an extensive correlation measure) depend on the mean number of event-wise low-$x$ gluons {\em cubed} suggesting a three-gluon interaction as the basic QCD mechanism. For arbitrary A-B collisions the trend vs $\bar \rho_0$ is determined for lower event multiplicities by the product $ \bar \rho_{sNN} \bar \rho_{hNN}$ with $N_{part} / 2 \approx N_{bin} \approx 1$ due to exclusivity and for higher multiplicities by the product $(N_{part} / 2) \times N_{bin}$ since  $ \bar \rho_{xNN}$ are slowly-varying for that condition. The general trend follows those elements over six orders of magnitude and unifies  systems from \pp\ to central \aa.
 
\subsection{Color dipole (dijet) angular correlations}

\begin{figure}[h]
\includegraphics[width=1.65in]{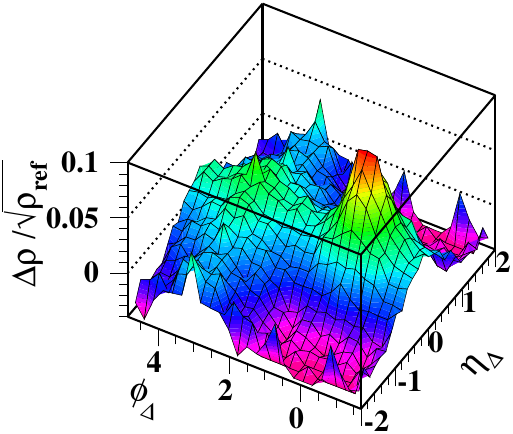}
\includegraphics[width=1.65in]{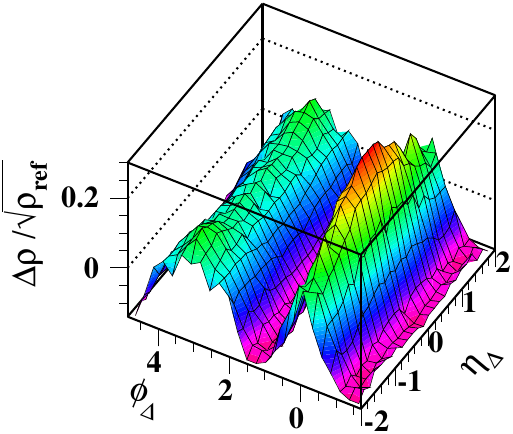}
	\caption{\label{aajetcorrs}
2D angular correlations for (left) most peripheral and (right) most central 200 GeV \auau\ collisions from Ref.~\cite{anomalous}. Compare the left panel to Fig.~\ref{ppcorrx} panels (c,d).
	}  
\end{figure}

Figure~\ref{aajetcorrs} shows angular correlations for the most peripheral (left) and most central (right) event classes from 200 GeV \auau\ collisions as reported in Ref.~\cite{anomalous}. Jet angular correlations are formally as they appear in Fig.~\ref{ppcorrx} (d): a same-side 2D peak at the origin representing {\em intra}jet correlations and an away-side dipole $\cos(\phi_\Delta - \pi)$ representing {\em inter}jet (jet-jet) correlations. Note that the left panel shows a superposition of soft (c) and hard (d) correlation shapes from Fig.~\ref{ppcorrx}. Accurate model fits to such general correlation data are required to isolate jet-related and non-jet (e.g.\ quadrupole) properties.

\begin{figure}[h]
	\includegraphics[width=1.65in]{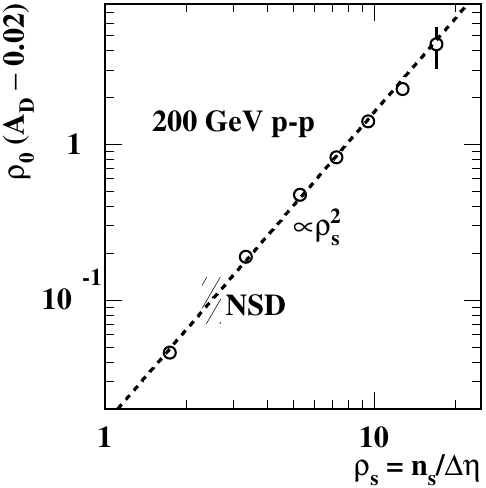}
\includegraphics[width=1.65in]{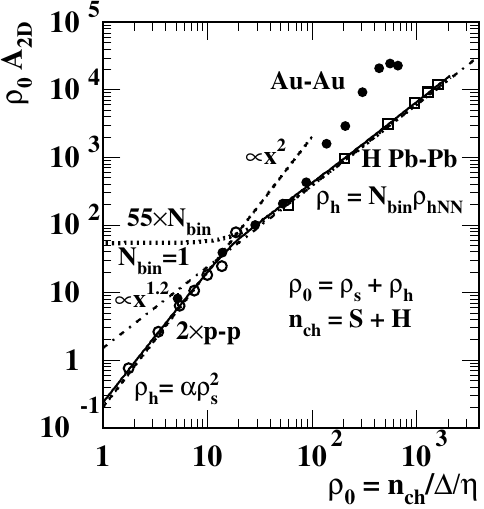}
	\caption{\label{aajetcorr}
	Left: Total correlated pairs for 200 GeV \pp\ away-side dipole 1D jet peak $\bar \rho_0 A_D$ (open circles) and observed quadratic trend (dashed).
	Right:  Dijet amplitude $\bar \rho_0 A_{2D}$ from 200 GeV \auau\ collisions (solid dots) and 2.76 TeV \pbpb\ collisions (solid squares)  vs charge density $\bar \rho_0$. \pp\ data and trend are copied from the left panel.
The solid curve is Eq.~(\ref{v22bxx}) rescaled to best match \pp\ and \auau\ data at lower \nch. {The bold dotted curve $\propto N_{bin}$ is discussed in the text.}
	} 
\end{figure}

Figure~\ref{aajetcorr} (left) shows the amplitude of the away-side dipole 1D jet peak (pair numbers) $\bar \rho_0 A_\text D$ (open circles, analogous to $V_2^2$) for 200 GeV \pp\ collisions from Ref.~\cite{ppquad}. 
The \pp\ dipole amplitude increases $\propto \bar \rho_s^2$ as expected from single-particle (SP) spectrum analysis with $\bar \rho_h \propto \bar \rho_s^2$~\cite{ppprd}, consistent with dijets as color dipoles resulting from two-gluon interactions. 
By analogy with Sec.~\ref{quadcorr} one may conjecture that dijet trends for \aa\ collisions should then vary as $\bar \rho_h \rightarrow N_{bin} \bar \rho_{hNN}$ with $\bar \rho_{hNN} \approx \alpha \bar \rho_{sNN}^2$. The corresponding equivalent to Eq.~(\ref{v22b}) would then be
\bea \label{v22bxx}
\bar \rho_0 A_\text{2D}(n_{ch}) &\propto& \bar \rho_h \propto  N_{bin} \ \bar \rho_{sNN}^2,
\eea
where $\bar \rho_0 A_\text{2D}$ is the same-side 2D jet peak amplitude.

Figure~\ref{aajetcorr} (right) shows quantity $\bar \rho_0 A_\text{2D}$ reflecting {\em total number of correlated pairs} as an { extensive} correlation measure (see beginning of Sec.~\ref{angcorrr}). That quantity corresponds to measure $\bar \rho_0 A_\text D$ in the left panel, both representing dijet production. The solid curve represents density $\bar \rho_h$ defined by Eq.~(\ref{v22bxx}) with parameters derived from App.~\ref{geometry}. Solid dots are 200 GeV \auau\ data from Ref.~\cite{anomalous}.  Open squares are integrated hard-component yields $H$ obtained from 2.76 TeV \pbpb\ $\pi$ spectra reported in Ref.~\cite{alicepbpb} rescaled by a single factor to match the solid curve. 
The \pp\ dashed line and open circles are from the left panel.
Measured jet amplitudes increase more  than four orders of magnitude across A-B systems. 

At lower \nch,  below a transition point near $\bar \rho_0 \approx \bar \rho_{s0} = 15$ noted for Fig.~\ref{ppquadjet} (right), the solid curve follows the \pp\ trend $\propto \bar \rho_s^2 \sim \bar \rho_0^2$. Above the transition the trend is $\propto N_{bin} \bar \rho_{sNN}^2$ where $\bar \rho_{sNN}$ for \aa\ collisions is observed to be approximately constant (see Fig.~\ref{story} (c)). Quantity $N_{bin}$ from App.~\ref{geometry} is shown as the bold dotted curve rescaled to match the solid curve above the transition. In that interval  $N_{bin} \propto \bar \rho_0^{1.2}$ as represented by the dash-dotted line.  Below the transition $N_{bin} \approx 1$ due to exclusivity.

\subsection{Conclusions re angular correlations}

The solid curve in Fig.~\ref{ppquadjet}  demonstrates that quadrupole data in the form $\bar \rho_0^2 v_2^2$ measuring  correlated pair numbers are proportional to a trend $\bar \rho_s \bar \rho_h$ over six orders of magnitude, where $\bar \rho_s$ and $\bar \rho_h$ are determined independently via SP spectra and \mmpt\ data.

The solid curve in Fig.~\ref{aajetcorr} is also based on a description of SP spectra (see Fig.~\ref{story} and Ref.~\cite{tompbpb}). Both {\em jet-related} trends provide accurate descriptions (modulo a single proportionality constant) of correlated-pair numbers below a transition point near $\bar \rho_0 \approx 100$. 
For \auau\ correlated-pair data presented in Ref.~\cite{anomalous} a {\em sharp transition} in jet properties (unusual amplitude increase) is noted. That novel result seemed to contradict expectations for jet ``quenching.'' A recent study~\cite{tompbpb} suggests that any reduction {\em at high \pt} relative to a linear model of jet fragment distributions arises from ``exclusivity''~\cite{tomexclude} and parton time dilation. No dense medium is evident.

It is notable that in relation to two distinct phenomena both correlation amplitudes (number of correlated pairs) and SP yields  follow the same trends derived from SP spectrum analysis over a range of systems from isolated \pp\ to central \aa. 
That congruence is consistent with hadron production arising from underlying mechanisms that produce isolated particles {\em and} correlated pairs both proportional to a number of {\em elementary processes} that varies with collision conditions. Thus, jet- or color-dipole-related hadron number {\em and} correlated pair number are both proportional to a number of gluon-gluon  collisions $\propto \bar \rho_h$. The same applies to color-quadrupole hadron production and three-gluon collisions $\propto \bar \rho_s \bar \rho_h$. It is difficult to explain those detailed trends based on hadron emission from a dense thermalized bulk medium undergoing large-scale fluid motion, also contradicted by survival of low-energy ($\approx$ 3  GeV) jets.

\section{Discussion} \label{disc}

Following the discovery of nucleon collectivity in Bevalac A-B collisions there has been a persistent tendency within the heavy ion community to interpret {\em apparently} similar phenomena in high-energy nuclear collisions preferentially as indicating QGP formation. Jet manifestations tend to be minimized and/or misinterpreted as part of an ongoing {\em confirmation bias}. This section considers that conflict. The material includes responses to QGP claims appearing in several review articles as in Sec.~\ref{promise}.

\subsection{Collectivity and flows in nuclear collisions} \label{}

Plastic flow of intact nucleons was reported at the Bevalac in 1984~\cite{poskanzer}. Collective flow, buttressed by arguments from astrophysicists studying neutron stars, was extrapolated to possible formation of a deconfined quark-gluon plasma at higher collision energies with  presumed higher temperatures and pressures~\cite{baym}. Such arguments motivated RHIC construction. Flow or {\em collectivity} was seen as a signature manifestation of QGP formation.

Since RHIC startup various strategies have been introduced to prefer flow-centric interpretations of data features over jet interpretations. 
Regarding SP spectra the \pt\ acceptance is conventionally divided into three regions: soft (0-3 GeV/c), hard (above 5-6 GeV/c) and intermediate with physical properties still debated. For example, see the spectrum description in Ref.~\cite{starspec04} wherein the boundaries are 2 GeV/c and 6 GeV/c. Monolithic (single-component) spectrum models such as blast-wave or Tsallis models are applied to a low-\pt\ fraction of the \pt\ acceptance, and model parameters are interpreted as representing thermodynamic properties -- temperature and radial flow of a dense medium. Variation of ensemble-mean \mmpt\ as a spectrum measure is described as ``an {\em unambiguous prediction} of the hydrodynamic framework of heavy-ion collisions [emphasis added]''~\cite{gardim}.

Such monolithic spectrum analysis conflicts with data spectra that exhibit distinct jet contributions (Sec.~\ref{structure}) and a two-component model (TCM) that describes spectra over large \pt\ intervals (0 to 20 GeV/c or more) within uncertainties (Sec.~\ref{heavyspectra}) as noted in the present study.

Regarding angular correlations ``...anisotropies in transverse-momentum distributions provide an {\em unambiguous signature} of transverse collective flow in ultrarelativistic nucleus-nucleus collisions [emphasis added]''~\cite{ollitrault}.
Angular correlations are separated into ``jet-like'' and ``flow-like'' components. Jet-like correlations are conventionally defined by application of  ``trigger-associated'' cuts wherein the trigger condition is typically $p_t > 4$ GeV/c (\yt\ $>$ 4). Such  conditions, presumably motivated by assumptions that jets are a ``high-\pt\ phenomenon,'' eliminate {\em almost all jet fragments} from resulting correlation data (see Fig.~\ref{ppcorrx}, b).
Surviving ``jet-like'' correlations then suffer from strong biases with ill-defined physics consequences.
``Flow-like'' (elliptic flow)  correlations are conventionally quantified as Fourier amplitudes ($v_2$ etc.) of correlated pairs projected onto 1D azimuth. Such projections may include jet contributions (``nonflow'') as substantial systematic biases.

With claimed observation of flow-like correlations in small systems comes the paradox of observing phenomena persisting in \pp\ collisions that were assumed unique for central \aa\ collisions. ``QGP droplets'' are said to appear even in \pp\ collisions. What is the detailed formation mechanism?  How are sufficient pressure gradients then developed to drive hydro flows? Where is the expected phase transition to a quark-gluon plasma? Why is jet quenching not observed in small systems as a result?

Hydro-based spectrum fits below 3 GeV/c are in effect deceptive. The same \pt\ interval contains 95\% of jet fragments with a peak mode near 1 GeV/c. So-called ``hardening'' of spectra results not from hydrodynamic flows but instead from increased jet production with increasing event multiplicity that is to be expected. There is no actual {\em boost} of the spectrum hard (jet) component on \yt\ that might arise from ``flow'' of a dense medium.

Theoretical arguments supporting viscous-hydro flows and ``jet quenching'' in small systems are inconsistent with a body of evidence from spectra and two-particle correlations. ``Flow-like features'' are simply explained by conventional QCD mechanisms properly understood. Jet fragment production over the full jet energy spectrum down to a few GeV should be properly acknowledged.

\subsection{Jet manifestations in nuclear collisions}

Jets from \ee\ collisions were observed in the mid seventies at SLAC~\cite{jethistory}. Studies at the CERN S$p \bar p$S in the mid eighties revealed jets from \ppbar\ collisions as definitive manifestations of QCD in hadronic interactions. High-\pt\ jets were acknowledged as distinct peaks in UA1 ``Lego plots'' on $(\eta,\phi)$~\cite{ua1jets2}. From such data it was recognized that a minimum-bias \ppbar\ jet energy spectrum might extend down to  $\approx 3$ GeV~\cite{minijets}.
From about 1990 to mid 2000s LEP and HERA measured the properties of monojets and dijets from isolated \ep\ and \ee\ events. QCD details governing parton fragmentation to jets were measured precisely. Collaborations at FNAL  measured dijet properties in \ppbar\ collisions. 
Yet substantial ignorance of previous high-energy physics (HEP) jet measurements seems commonplace within the heavy ion community.

With commencement of RHIC operations an {\em Event Structure} working group within the STAR collaboration undertook studies  of jet properties in \pp\ and \auau\ collisions, specifically two-particle correlations and spectrum structure. A  two-component (soft+hard) spectrum model emerged from \pp\ spectrum studies in 2003~\cite{ppprd}. Detailed studies of \pp\ 2D angular correlations and \yt-\yt\ correlations were completed by 2005~\cite{porter2,porter3}. Compact parametrizations of \ee\ dijet fragmentation functions and \pp\ jet energy spectra were reported in Refs.~\cite{eeprd,jetspec2}. By convoluting those {\em measured} jet energy spectra with {\em measured} fragmentation-function ensembles TCM spectrum hard components for \pp\ and \aa\ collisions were {\em predicted quantitatively} down to \pt\ $\approx$ 0.35 GeV/c~\cite{fragevo}.

Since 2010 a series of studies of \pp, \pa\ and \aa\ collision systems has provided important analysis infrastructure and led to new insights about high energy collisions, including the central role of quantum mechanics and relativistic time dilation. Details include
Ref.~\cite{jetspec} where jet fragment yields are predicted from measured jet correlations,
Ref.~\cite{jetspec2} where several aspects of jet formation are considered comprehensively,
Ref.~\cite{ppquad} where \pp\ angular correlations and SP densities reveal jet {\em and quadrupole} properties as a fundamental reference,
Ref.~\cite{alicetomspec} where an extended study of \pp\  \pt\ spectra also provides a fundamental reference
and
Ref.~\cite{mbdijets} where minimum-bias dijets and nuclear transparency in various systems are subjects of a broad study.

Reference~\cite{tomglauber} demonstrates that the classical Glauber model as applied to \ppb\ spectrum data is strongly biased but that accurate collision geometry may be inferred from ensemble \mmpt\ data. 
Reference~\cite{tomexclude} reports discovery of {\em exclusivity} for \nn\ collisions and its relation to the Glauber model. 
Reference~\cite{njquad} considers methods to isolate precisely a {\em nonjet} quadrupole from jet structures.
Reference~\cite{ppbpid} reports an accurate spectrum model for identified-hadron (PID) spectra from \ppb\ collisions based on exclusivity and  exact solution of \ppb\ geometry.

Reference~\cite{tomnewppspec} reports an extensive study of nonPID aspects of \pp\ spectra including biases from event selection.
Refences~\cite{pidpart1,pidpart2} present an extended precision analysis of PID \ppb\ spectra as followups to Ref.~\cite{ppbpid}.
Reference~\cite{pppid} reports a broad analysis of \pp\ PID spectra.
Reference~\cite{ppbnmf} considers the presence (or not) of ``jet quenching'' in \ppb\ collisions as a small collision system.
Reference~\cite{tompbpb} considers the role of quantum transitions and time dilation in A-B collisions and a novel approach to the question of jet quenching.
Reference~\cite{tomnewquad} presents a new analysis of quadrupole phenomenon including quadrupole spectra and power-law trends for quadrupole amplitudes vs charge density $\bar \rho_0$.

The cited publications lay out a broad array of detailed quantitative evidence for manifestations of minimum-bias jet production in spectra and two-particle correlations that is compatible with QCD principles. Hadron production arises almost entirely from two basic fragmentation processes: projectile-nucleon dissociation and splitting cascades from scattered partons. The system of observations conflicts with descriptions based on flowing bulk matter described by hydro and an equation of state.

A comprehensive alternative description has emerged recently based on the concept of isolated few-gluon interactions~\cite{tompbpb,tomnewquad}. The description applies consistently to color quadrupole ($v_2$) and color dipole (dijets) production across all A-B collisions from \pp\ to central \aa. That is a novel subject of Sec.~\ref{2dcorr} which remains consistent with previous results from TCM spectrum analysis.

\subsection{Responses to some prominent review articles}

Certain published review articles, conventionally considered representative of the heavy-ion community, assert strong conclusions about the results of theoretical and experimental research by that community, in particular about production of a quark-gluon plasma and its properties. Careful comparison of alternative research results produced over the past twenty years with certain details of the review articles suggests that at least some claims are unjustified and that confirmation bias may play a  role. In what follows three such reviews are considered.

\subsubsection{Perfect fluid at the RHIC}

The following text includes responses to statements in Sec.~\ref{perfect} from Ref.~\cite{perfect} (2004).  Quotes are italicized.

{\em ...the flow pattern of thousands of produced hadrons is the primary observable used to look for novel collective phenomena.} Presumably ``flow pattern'' refers to statistical measure $v_2$ conventionally associated with elliptic flow. As demonstrated in this study the same collision processes persist from lowest-\nch\ \pp\ to central \aa, with correlated pairs for quadrupole structure increasing over six orders of magnitude, from a few in a fraction of \pp\ events to possibly hundreds in any central \aa\ event. Thus, the ``primary observable'' cannot be assigned  to a particular property or state of a class of nuclear collisions.

{\em ...collective flow properties test  two of the conditions necessary for the validity of QGP....the extent of thermalization and the equation of state \rm{[EoS]}.}
Survival of low-energy jets and stability of spectrum soft components vs \aa\ geometry variation argue against thermalization via particle rescattering. There is no indication of a phase transition (interruption of data trends) over a broad range of collision conditions, hence no evidence for an EoS relating to a thermodynamic state of matter.

{\em Elliptic flow measurements confirm that} \rm{[QGP]} {\em is...in local thermal equilibrium....} This apparently refers to comparison of hydro theory output and $v_2(p_t)$ data. {\em Predictivity} of complex hydro Monte Carlos as distinct from an elaborate fitting exercise may be questioned.

{\em ...the hadron mass dependence of the flow pattern is remarkably consistent with...QCD...computations of the equation of state.} As demonstrated in Ref.~\cite{tomnewquad}  the mass dependence of $v_2(p_t)$ is consistent with a narrow range of boost values that does not correspond to Hubble-like expansion of a bulk medium as expected for \aa\ hydro.

{\em Theoretical analysis of jet quenching confirm \rm{[sic]} {\em the energy density estimated. They give large energy losses for jets through the matter...and strengthen the case for multiple strong interactions of the quark and gluon constituents....} This refers to results from rescaled spectrum ratio $R_\text{AA}$. But $R_\text{AA}$ suppresses almost all jet contributions at lower \pt\ and is rescaled with a factor $N_{bin}$ that is strongly biased. Analysis of PID spectra from a range of A-B systems reveals that jet contributions at lower \pt\ exhibit no significant attenuation with varying \nch\ or centrality. Variations at higher \pt\ are consistent with effects of exclusivity and time dilation as in Ref.~\cite{tompbpb}.

{\em Criteria for the discovery of QGP at RHIC:
(a) matter at energy densities so large that quarks and gluons are the degrees of freedom,
(b) the matter is thermalized,
(c) properties of hot and dense matter must follow from QCD computations based on hydrodynamics, LGT\footnote{lattice gauge theory} and pQCD for...jets. All of the above are satisfied from the published data at RHIC.} It is important to note that one of the experimental white papers, Ref.~\cite{starwhite} (STAR, 2005, 4486), includes  the following disclaimer in its abstract: ``However, the measurements themselves do not yet establish unequivocal evidence for a transition to this new form of matter [QGP].''
Subsequent research, such as that reviewed in this study, includes the following results: Based on \mmpt\ data the density of participant nucleons is typically one third of Glauber Monte Carlo estimates~\cite{tomglauber}. The TCM soft component, experimentally what dominates hadron production, shows no indication of interacting with the hard component and is typically independent of varying collision conditions. The hard component (jets) shows no alteration by passage through a dense medium and is consistent with measured properties of \pp\ jets as to parton energy spectra and fragmentation functions. Thus, the stated criteria are not satisfied.

\subsubsection{Dense medium in small systems I}

The following text includes responses to statements in Sec.~\ref{small1} from Ref.~\cite{buzsa}. A primary concern  is underestimation of jet production in high-energy nuclear collisions.

Concerning jet production [a] {\em small fraction of incident partons suffer\rm [s] \em hard perturbative interactions...which...lead to a relatively improbable...production of particles with high transverse momentum.} That is a major misrepresentation of jet production in high-energy collisions. For 200 GeV \pp\ collisions the jet probability for NSD collisions (minimum-bias average) may be a few percent, but for \pp\ collisions selected  for charge densities ten-fold greater several jets per collision may be produced. For central \pbpb\ collisions several tens of percent of produced hadrons may be jet fragments where the fragment distribution extends down to 0.5 GeV/c.

{\em Heavy ion collisions quickly form a droplet of quark gluon plasma (QGP) with remarkably small viscosity.} The term ``droplet'' suggests liquid consistent with a  ``perfect fluid'' conjecture~\cite{perfect}. However, multiple data manifestations are inconsistent with QGP formation.

{\em They {\rm [partons] {\em are so strongly coupled...that they form a collective medium that expands and cools...with a  remarkably low viscosity-to-entropy-density ratio....} The statement again reflects a preferred narrative, but evidence for a dense expanding medium is not apparent.

{\em What did come as a surprise is how many other phenomena are similar in AA and pA collisions and even in pp collisions....} A central premise for the RHIC program was assumption that small collision systems should serve as {\em references} for QGP formation in \aa\ collisions: If phenomena ``are similar'' in small and large systems the {\em scientific} conclusion should be that there is no QGP formation in the latter given known properties of the former. The argument above appears as confirmation bias.

...including {\em the azimuthal anisotropies {\rm [elliptic flow] {\em encoded in multiparticle correlations that were once thought to be unique to AA collisions.} Again, observation of a quadrupole component in \pp\ collisions should prompt the conclusion that the phenomenon is not associated with formation of a dense QCD medium, is instead a new elementary interaction consistent with standard QCD.

{\em ...it is tempting to interpret them \rm [similarities of phenomena] {\em as indicating  that proton-size droplets of QGP can be formed in those pp and pA collisions....}   The temptation should be resisted as a form of confirmation bias.

{\em ...in occasional heavy ion collisions, partons from the incident nuclei scatter off each other with very large momentum transfer, creating two or more quarks or \rm{[gluons]} {\em with very high transverse momentum...(many tens of GeV at RHIC; as high as 100 or even 1000 GeV at the LHC).} Again, that is a misrepresentation of jet production. While rare events may follow that description the {\em vast majority of jets} appears near the lower end of the jet energy spectrum $\approx 3$ GeV. The corresponding mode of the jet {\em fragment} \pt\ distribution appears near 1 GeV/c.

{\em Jet production and showering in a vacuum is well described by perturbative QCD.} What is described by pQCD is parton scattering cross sections. Parton distribution functions (nucleon PDFs) and fragmentation functions (parton FFs) are measured but may be {\em approximated} by pQCD over limited intervals. Description of jet production requires a combination of those elements.

{\em When a jet is produced in a HI collision the partons must plow through the droplet of QGP produced in the same collision.} An array of evidence (as noted herein) demonstrates that there are no droplets and no plowing. Jet production in \pa\ and \aa\ collisions also involves quantum transitions and relativity, with major consequences for fragment yields vs \pt\ as demonstrated in Ref.~\cite{tompbpb}. A holistic and quantitative description of collision dynamics, including both quadrupole and dijet production as presented in this study, resolves seeming paradoxes within a context based on conventional QCD.

\subsubsection{Dense medium in small systems II}

This text includes responses to statements in Sec.~\ref{small2} from Ref.~\cite{nagle}. A general concern is confirmation bias wherein evidence  contradictory to  a preferred narrative motivates abandonment of basic physics principles.

{\em Crucial information regarding collectivity is garnered through the measurement of...particle correlations} on 2D angle space \rm [$(\eta_\Delta,\phi_\Delta)$]. Measured angular correlations do provide crucial information about jet production and a nonjet quadrupole. The term ``collectivity'' is not clearly defined, relating vaguely and generally to ``flows.''

Referring to jet production {\em these dijet correlations are subdominant for $p_t < 5$ GeV/c.} The musical term ``subdominant'' is here applied as a neologism to suggest that jet-related correlations are relatively negligible below 5 GeV/c. The claim that jets play no role below 5 GeV/c (are ``subdominant'') is demonstrably false, contradicted by yield, spectrum and correlation data. For example,  Fig.~\ref{ppcorrx} (b) shows a correlated-pair distribution on 2D transverse-rapidity space $(y_{t1},y_{t2})$ with hard-component peak centered near (2.7,2.7) [(1,1) GeV/c]. Figure~\ref{ppcorrx} (d) shows selected hadron pairs with mean values $p_t \approx 0.6$ GeV/c exhibiting a strong same-side 2D jet peak representing {\em intra}jet correlations. Clear jet structure for low-\pt\ hadron pairs in panel (d) is thus directly linked to the corresponding low-\pt\ spectrum hard component in panel (a).
This example demonstrates a trend to obscure jet contributions by reserving {\em all} correlations at lower \pt\ to collectivity ($\Rightarrow$ flows). Detailed analysis says otherwise. 


{\em It now may seem surprising that no jet quenching effect is apparent in p + A collisions if indeed a hot medium is formed.} There are no surprises if (a) there is no hot medium formed in \pa\ (or anywhere else) and (b) there is no real ``jet quenching' effect in any collision system.
Higher-\pt\ spectrum modifications conventionally interpreted to indicate jet quenching are detectable in \ppb\ collisions as well but have nothing to do with a hot medium. Reference~\cite{tompbpb} reports a recent study of 2.76 TeV \pp\ and \pbpb\ spectra and 5 TeV \ppb\ spectra that reveals the consequences of {\em exclusivity} and relativistic time dilation. Those are the underlying causes of variations in the high-\pt\ structure of A-B spectra relative to \pp\ spectra. They do not relate to a dense medium.

{\em In summary, ample theoretical arguments...suggest that viscous relativistic hydrodynamics can be applied to describe particle production and flow in p+p and p+A collisions at high energies.} According to the scientific method theoretical conjectures are subject to {\em skeptical challenge} via experimental tests. Many (``ample'') such conjectures, especially in the absence of {\em rigorous} experimental tests, cannot validate preferred conclusions.

{\em ...the field of relativistic heavy ion physics is in the midst of a revolution in our understanding of the conditions necessary for...near-perfect fluid...described by viscous relativistic hydrodynamics.} {\em The revolution...[is] driven by  the experimental observation of flow-like features in the collisions of small hadronic systems.} This is an archetypal example of confirmation bias. The desired outcome is a ``near-perfect fluid'' and ``''viscous hydrodynamics'' as claimed for RHIC results c.\ 2005. The observation of ``flow-like features'' is based on misinterpretation of data structure more simply explained by conventional QCD that persists within all collision systems.

One may note that the term ``propaganda'' emerged (c.\ 1620) to express Catholic church interest in ``propagation of the faith'' (i.e.\ Catholic religion) across populations. In more recent times propaganda has been manifested as biased presentations that exaggerate or even fabricate claims to promote a viewpoint. Propaganda is often employed by large organizations to shape public opinion.

\section{Summary}\label{summ}

This article responds in part to claims of quark-gluon plasma (QGP) formation in small collision systems (e.g., \pp, \ppb) motivated by certain results from the  large hadron collider (LHC) during the past fifteen years wherein data features attributed previously to manifestations of elliptic flow and jet quenching in \auau\ collisions at the relativistic heavy ion collider (RHIC) appeared also at significant levels in smaller systems at the LHC.

Experimental evidence accumulated over more than ten years strongly indicates that a popular paradigm assigned to high-energy collisions in a heavy-ion context is not valid. Conventional assumptions that jets do not contribute significantly to hadron production below a transverse momentum value $p_t \approx 5$ GeV/c are demonstrably false as shown in Sec.~\ref{structure}. Assumptions that collision systems are in some sense dense conflict with direct determination of participant nucleon densities via {\em measured} ensemble-mean \mmpt\ data as in App.~\ref{geometry}.

Measurement of statistical quantity $v_2$ (interpreted to represent ``elliptic flow'' of a hot and dense flowing medium) is defined as the square root of a pair ratio. Its denominator is an observed SP spectrum whose structure, as described in Sec.~\ref{structure}, is clearly dominated by jet contributions. Its numerator is a Fourier amplitude for 2D angular correlations on  $(\eta,\phi)$. In previous studies Fourier amplitudes, {\em not} in ratio, were revealed as {\em quadrupole spectra} well approximated by a Boltzmann exponential on transverse mass $m_t$ {\em in a boost frame} with low slope parameter $T\approx 95$ MeV compared to SP spectrum soft component $T \approx 150$-200 MeV. The same fixed quadrupole model describes several hadron species in common. When back transformed that model describes $v_2(p_t)$ data precisely from \pt\ = 0 to 6.5 GeV/c or higher.

Recent studies of two-particle angular correlation measurements obtained over more than twenty years reveal a major simplification of collision models. 2D angular correlations from 200 GeV \pp\ collisions measured in 2015 revealed that a quadrupole component with amplitude measured by number of correlated pairs (extensive measure) increases as the {\em cube} of the charge-density soft component (arising from low-$x$ gluons). That trend suggested that the quadrupole arises from three-gluon interactions.

Studies of hadron production in \pp\ and \ppb\ collisions have led to dramatic improvement in A-B geometry determination, not by a classical Glauber Monte Carlo model but by inversion of ensemble-mean \mmpt\ data. Results include the discovery of {\em exclusivity} wherein a nucleon in an A-B collision may interact with only one other nucleon {\em at a time}. As a consequence \aa\ collisions are restricted to single \nn\ collisions over a substantial event-charge \nch\ lower interval as described in App.~\ref{geometry}.

Combining quadrupole results from \pp\ collisions with the revised A-B geometry determination including exclusivity has resulted in a general relation that predicts variation of quadrupole amplitudes (measured as pair numbers) with event multiplicity \nch\ for any \mbox{A-B} collision system as summarized  in Sec.~\ref{2dcorr}. The same approach applied to jet-related correlations leads to a predicted {\em quadratic} trend versus \nch\ appropriate for two-gluon interactions. One may thus describe all A-B  systems in terms of few-gluon interactions and  color multipole radiation: dipoles (dijets) for two-gluon interactions and quadrupoles for three-gluon interactions, recalling that momentum conservation for a three-body interaction is quite different from that for a two-body interaction.

From those developments one may conclude that high-energy A-B collisions  are low density, and hadron production is dominated by three processes: (a) participant-nucleon dissociation (spectrum soft components), (b) two-gluon interactions leading to color dipole radiation or dijets (spectrum hard components) and (c) three-gluon interactions leading to color quadrupole radiation (not yet identified within SP spectra). Processes (a-c) are clearly identified within 2D angular correlations for all systems. Given  these recent results it is unlikely that a ``hot and dense QCD medium'' is produced or that hydrodynamic theory is relevant for any system.

\begin{appendix}

\section{TCM for A-B PID spectra} \label{tcmpid}

The TCM has been derived from spectrum data for a variety of A-B collision systems by what has been described by some as a {\em mere fitting procedure}. The process is correctly described as a form of {\em lossless data compression} in which a large volume of particle data is reduced to a few parameter values by a process that {\em does not rely on model assumptions}. Derivation of a TCM from 200 GeV \pp\ \pt\ spectra was first described in Ref.~\cite{ppprd}. Some relevant details are illustrated by similar analysis of 13 TeV \pp\ spectra and 5 TeV \ppb\ spectra in Sec.~\ref{structure}. The overall model depends almost entirely on accurate determination of jet contributions to spectra.

In what follows monopole spectrum $\bar \rho_0(x_t)$ is a density on variable $x_t$ ($p_t$ or $y_t$). $\bar \rho_2(x_t)$ denotes {\em  quadrupole} spectra as described in Refs.~\cite{quadspec,tomnewquad}.
The TCM for unidentified-particle (nonPID) data spectra has the form
\bea \label{tcm}
\bar \rho_0(p_t,n_{ch}) &=& \bar \rho_s(n_{ch}) \hat S_0(p_t) + \bar \rho_h(n_{ch}) \hat H_0(p_t),
\eea
where $\bar \rho_x(n_{ch})$ is a particle density on pseudorapidity $\eta$ or longitudinal rapidity $y_z$ with charged-particle number $n_{ch}$ serving as  event-class index. $\hat X_0$ are unit-integral model functions for soft (s, S) and hard (h, H) model components. Each component is then factorizable into \nch-dependent and \pt-dependent factors. Model functions $\hat X_0(p_t)$ are inferred from data as shown below.

For \pp\ collisions the relation $\bar \rho_h \approx \alpha \bar \rho_s^2$ is observed to hold accurately~\cite{ppprd}, where $\alpha \approx O(0.01)$ varies slowly with collision energy (see Ref.~\cite{alicetomspec}, Fig.~16, left).
For \mbox{A-B} collisions $\bar \rho_s = (N_{part}/2)\bar \rho_{sNN}$ and $\bar \rho_h =N_{bin} \bar \rho_{hNN}$, the factors representing number of nucleon N participant pairs and binary \nn\ collisions while $\bar \rho_{xNN}$ are densities {\em per participant pair} averaged over \aa\ collisions. 

{\em Hard/soft density ratio} $x\nu \equiv \bar \rho_h/\bar \rho_s$, where  $x \equiv \bar \rho_{hNN} / \bar \rho_{sNN}$ and $\nu \equiv 2 N_{bin} / N_{part}$ with $x \approx \alpha \bar \rho_s$ for \pp\ collisions,
is a fundamental TCM parameter representing the ``centrality'' of \aa\ collisions inferred from ensemble-mean \mmpt\ data~\cite{ppbpid,pidpart1}, not a classical Glauber Monte Carlo~\cite{tomglauber,tomexclude}. A-B impact parameters and overlap areas seem not critical to collision dynamics.

A PID TCM for hadron species $i$ is obtained by generalizing Eq.~(\ref{tcm}) as
\bea \label{pidtcmm}
\bar \rho_{0i}(p_t,n_{ch}) \hspace{-.04in} &=&\hspace{-.04in} \bar \rho_{si}(n_{ch}) \hat S_{0i}(p_t) + \bar \rho_{hi}(n_{ch}) \hat H_{0i}(p_t),~~~~~~
\eea
where $\bar \rho_{si} = z_{si}(n_{ch}) \bar \rho_{s}$ and $\bar \rho_{hi} =z_{hi}(n_{ch}) \bar \rho_{h}$.
The $z_{xi}(n_{ch})$ factors are specific soft and hard {\em fractional abundances} $\leq 1$. The relation $\bar \rho_{0i}(n_{ch})  = z_{0i}(n_{ch}) \bar \rho_0(n_{ch})$ for particle densities defines $z_{0i}$ as {\em total} fractional abundances related to statistical models~\cite{statmodel}. Given those details $z_{si}(n_{ch})$ may be defined as
\bea \label{zxi}
z_{si}(n_{ch})&=& \frac{1+x \nu}{1+\tilde z_i x \nu} z_{0i}(n_{ch}),
\eea
with $\tilde z_i \equiv  z_{hi} / z_{si}$. $z_{si}(n_{ch})$ and $z_{hi}(n_{ch})$ values have been inferred from 5 TeV \ppb\ spectra in Ref.~\cite{pidpart1}. 

Figure~\ref{tildezparamss} (left) shows ratios $\tilde z_{i}(n_s)$ (points) inferred from $z_{si}(n_s)$ and $z_{hi}(n_s)$ data reported in Ref.~\cite{pidpart1} for charged hadrons (solid dots) and neutral hadrons (open circles), from $\pi$s at the bottom to $\Lambda$s at the top. Dashed and dotted lines corresponding to kaons and baryons are best fits by eye to the points with Eq.~(\ref{tildezz}) 
\bea \label{tildezz}
\tilde z_i(n_s) &=& \tilde z_i^* + \delta \tilde z_i^* x(n_s)\nu(n_s).
\eea
Parameters are presented as solid dots in the right panel.
From those data it is apparent that parameter $\tilde z_i$ is simply proportional to hadron mass. The two proportionality constants then serve to generate all $z_{xi}(n_{ch})$ values from Eq.~(\ref{zxi}) with $x\nu$ from  nonPID TCM Fig.~(\ref{inverse}). Note that for $x\nu \rightarrow 0$ measured $z_{si}(n_{ch})$ values extrapolate to $z_{0i}$, thus estimating that parameter as the limiting case.

\begin{figure}[h]
	\includegraphics[width=3.3in]{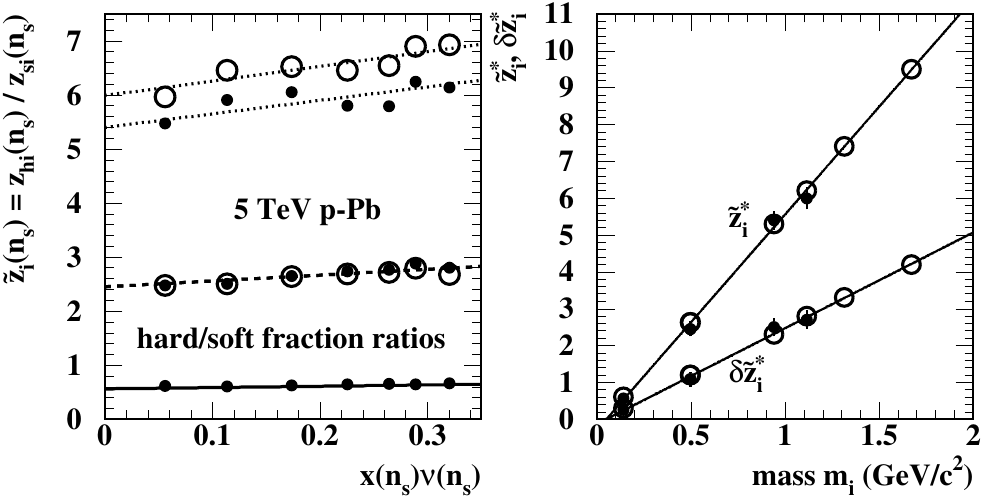}
	\caption{\label{tildezparamss}
		Left: Ratios $\tilde z_{i}(n_s) = z_{hi}/z_{si}$ inferred from $z_{si}$ and $z_{hi}$ measurements reported in Ref.~\cite{pidpart1} for charged (solid dots) and neutral (open circles) hadrons. The lines are linear parametrizations $\tilde z_i = \tilde z_i^* + \delta \tilde z_i^* x(n_s) \nu(n_s)$ that describe the ratio data: solid, dashed and dotted for $\pi$s, kaons and baryons respectively.
		Right: Coefficients $\tilde z_i^*$ and $\delta \tilde z_i^*$ for linear descriptions of ratio data in the left panel plotted vs hadron mass. The lines represent proportionality to hadron mass. The solid dots are values inferred in Ref.~\cite{pidpart1}. Open circles provide linear extrapolations on hadron mass for $\Xi$ and $\Omega$.
	}  
\end{figure}

Figure~\ref{tildezparamss} (right) shows  $\tilde z_i^*$ and $\delta \tilde z_i^*$ (solid dots) for  Eq.~(\ref{tildezz}) plotted vs hadron mass. Starred $\tilde z_i(n_s)$ parameters in Eq.~(\ref{tildezz}) are simply proportional to hadron mass and show no sensitivity to hadron identity or strangeness. Empirical mass trends enable predictions (open circles) of spectra for more-massive hadrons (e.g.\ $\Xi$s and $\Omega$s) from lower-mass hadron spectra~\cite{ppsss}.

\section{Geometry model for A-B collisions} \label{geometry}

Accurate determination of A-B collision geometry (number of nucleon N participants $N_{part}$, \nn\ binary collisions $N_{bin}$, soft and hard particle densities $\bar \rho_s$ and $\bar \rho_h$ vs measured total charge density $\bar \rho_0$) is essential to understand hadron production manifestations. The determination relies on measurements of jet phenomena.
A study of Glauber centrality for 5 TeV \ppb\ collisions~\cite{tomglauber} reveals  that a Monte Carlo method assuming the eikonal approximation for \nn\ collisions returns much larger participant numbers than is consistent with \ppb\ \mmpt\ data. A strategy based on \mmpt\ data has been developed to estimate \pbpb\ geometry parameters.

Corrected (for low-\pt\ spectrum acceptance cutoff) \pbpb\ $\bar p_t$ data have the general TCM description
\bea \label{ptprime}
\bar p_t \equiv \frac{\bar P_t} {n_{ch}} &\approx & \frac{\bar P_{tsNN} + \nu(n_s) \bar P_{thNN}(n_s)}{n_{sNN}(n_s) + \nu(n_s) n_{hNN}(n_s)}~~~
\\ \nonumber
&\approx &  \frac{\bar p_{ts} + x(n_s)\, \nu(n_s) \bar p_{th}(n_s)}{1 + x(n_s) \nu(n_s)},
\eea
where $\bar P_{txNN}$ is the total soft or hard \pt\ per \nn\ pair with $\bar p_{tx} = \bar P_{txNN} / n_{xNN}$ averaged over \nn\ pairs.
Total integrated charge is $n_{ch} = n_s + n_h$. 
Details of \pt\ production may be simply isolated by forming the product 
\bea \label{proddd}
\frac{n_{ch} }{n_{s}}  \bar p_t(n_s) &\approx& \frac{\bar P_t}{n_{s}} \approx \bar p_{ts} + x(n_s)   \,  \nu(n_s)\, \bar p_{th}(n_s),~
\eea
where $\bar p_{tx}$ are derived from TCM model functions $\hat S_0$ and $\hat H_0$.
Jet production is measured by increase of \mmpt\ above $\bar p_{ts}$.
Equation (\ref{proddd}) is crucial because $n_s$ cancels in ratio and product $x(n_s)\nu(n_s)$ is then inferred from \mmpt\ data.  

\begin{figure}[h]
	\includegraphics[width=1.65in]{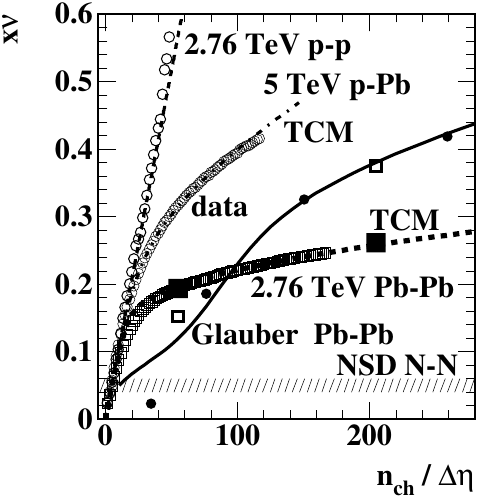}
	\includegraphics[width=1.65in]{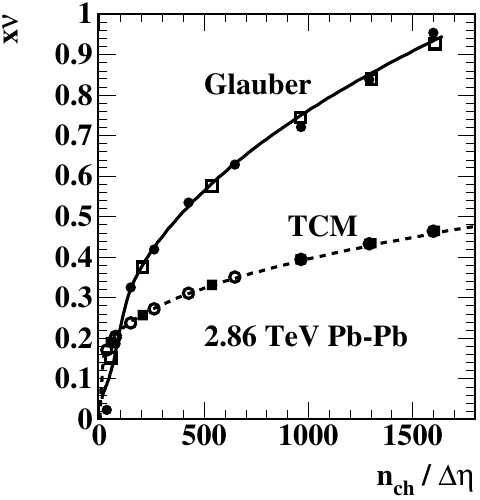}
	\caption{\label{inverse}
		Left: Product $x\nu$ (open points) extracted from \mmpt\ data via Eq.~(\ref{proddd}). Dash-dotted and upper dashed curve are TCM predictions. The lower dashed curve is a TCM description (ad hoc model) of \pbpb\ data. The solid points and curve are derived from Glauber results in Ref.~\cite{alicepbpbyields}. 
		Right: Extrapolation of the TCM trend from the left panel over the full \pbpb\ multiplicity range. The Glauber points and curve are  from Ref. \cite{alicepbpbyields}.
	}  
\end{figure}

Figure~\ref{inverse} (left) shows $x(n_s)\nu(n_s)$ (points) inferred by inverting \mmpt\ data from three collision systems -- \pp, \ppb\ and \pbpb. Curves for \pp\ and \ppb\ systems are TCM descriptions. 
\ppb\ \mmpt\ trends reveal competition for jet-related \mmpt\ increase between increasing participant N pairs measured by $x(n_s)$ and increasing \nn\ binary collisions measured by $\nu(n_s)$. \pbpb\ \mmpt\ results suggest that similar competition occurs in \pbpb\ for more-peripheral collisions, but factorization of $x\nu$ is not simple as  for \ppb.

Figure~\ref{inverse} (right) shows  Glauber model (solid dots and curve) and \pbpb\ $x\nu$ data (solid squares, dashed TCM extrapolation)  over the full range of \pbpb\ charge densities. 
The final TCM $x\nu$ values, 
plotted as  solid squares in Fig.~\ref{inverse}, indicate extrapolated model values corresponding to data $\bar \rho_0$ values, not separate measurements. 
These $x\nu$ estimates correspond to {\em integrated hard-component $\bar P_{th}$} in Eq.~(\ref{ptprime}), not differential fragment distributions on \pt\ or \yt. Below is a detailed description of the model. 

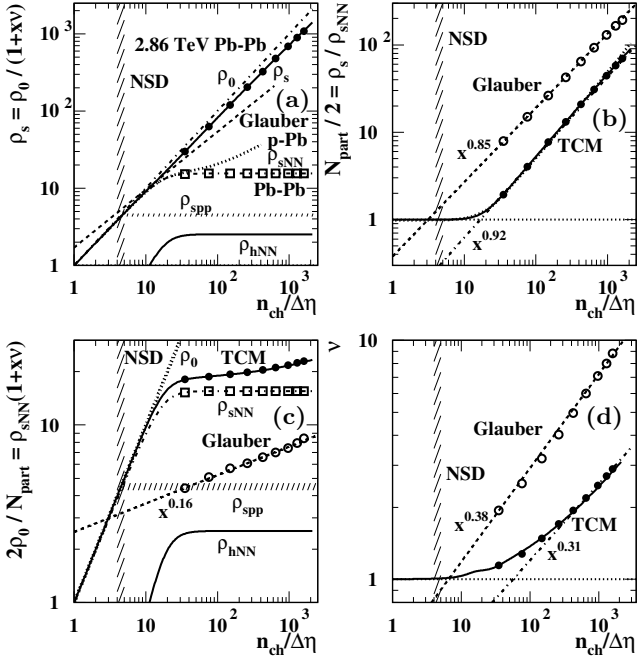
\begin{figure}[h]
	\includegraphics[width=3.3in]{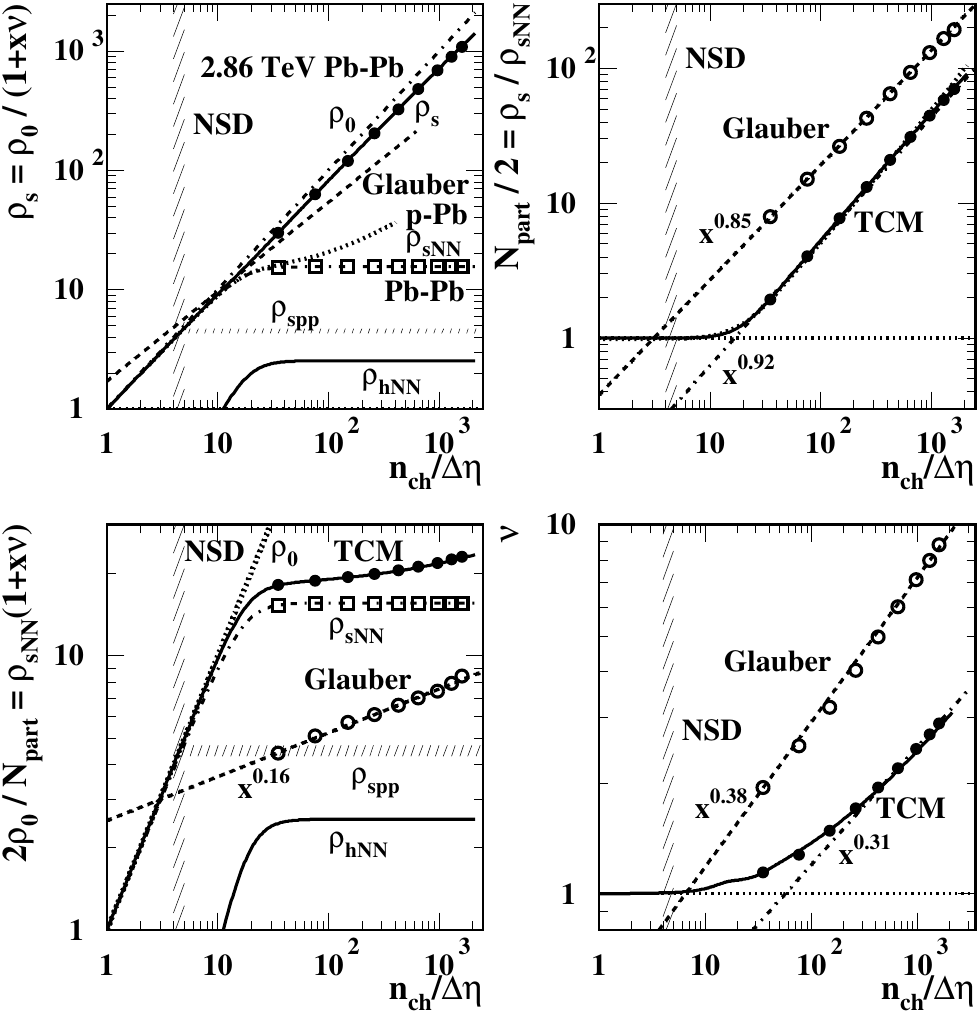}
\put(-137,207) {\bf (a)}
\put(-18,200) {\bf (b)}
\put(-138,88) {\bf (c)}
\put(-20,88) {\bf (d)} 
	\caption{\label{story}
		(a) Soft density $\bar \rho_s = \bar \rho_0 / (1+x\nu)$ (solid dots and curve) obtained with TCM $x\nu$ from Fig.~\ref{inverse}. The dash-dotted curve is a $\bar \rho_0$ reference. The dashed curve is obtained with Glauber $x\nu$ in  Fig.~\ref{inverse}. The lower dash-dotted and dotted curves are TCM $\bar \rho_{sNN}$ trends for \pbpb\ and \ppb\ collisions respectively.
		(b) $N_{part}/2 = \bar \rho_s / \bar \rho_{sNN}$ (TCM points and curve) derived from $\bar \rho_s$ (solid) and $x = \alpha  \bar \rho_{sNN}$ conjecture (lower dash-dotted) in (a). Open circles are Glauber (primed) values from Ref.~\cite{alicepbpbyields}.
		(c) $2 \bar \rho_0 / N_{part} = \bar \rho_{sNN} (1+x\nu)$ for TCM (solid points and curve) and Glauber model (open points and dashed curve)~\cite{alicepbpbyields}.
		(d) Quantity $\nu = 2 N_{bin}/N_{part}$ from the TCM (solid) and from the Glauber model (dashed). 
	}  
\end{figure}

Figure~\ref{story} (a) shows soft-component density $\bar \rho_s$ (points, solid curve) derived from measured density values $\bar \rho_0$ from Ref.~\cite{alicepbpbyields} (solid points) as $\bar \rho_s = \bar \rho_0/(1+x\nu)$ where product $x\nu$ is inferred from \mmpt\ data in Fig.~\ref{inverse} (solid squares, dashed curves). 
The dashed line shows  $\bar \rho_s$ estimated with product $x\nu$ from the Glauber model. 
The lower dash-dotted and dotted curves are $\bar \rho_{sNN}$ for \pbpb\ and \ppb\ data. The open boxes are estimated \pbpb\ $\bar \rho_{sNN}$ values. 
 The lower solid curve is a corresponding  trend for hard component $\bar \rho_{hNN} \approx \alpha \bar \rho_{sNN}^2$ as for \pp\ collisions~\cite{ppprd}.

Figure~\ref{story} (b) shows TCM estimates for $N_{part}/2 = \bar \rho_s / \bar \rho_{sNN}$ (solid points, solid curve) based on estimates of $\bar \rho_{sNN}$ shown in panel (a), (lower dash-dotted). Open circles are  Glauber estimates vs $\bar \rho_0$ from Ref.~\cite{alicepbpbyields}.

Figure~\ref{story} (c) shows TCM trend $2\bar \rho_0 / N_{part} = \bar \rho_{sNN} (1 + x\nu)$ vs $\bar \rho_0$ (solid dots and curve). Also shown is the result from Glauber analysis (open circles, dashed) reported in Ref.~\cite{alicepbpbyields}.
The dash-dotted curve is an estimated trend for quantity $\bar \rho_{sNN}$ from (a).
In effect, $\bar \rho_{sNN} = \bar \rho_s$ ($\nu = 1$) up to a transition point $\bar \rho_{s0} \approx 15$ and then remains approximately constant for more-central \aa\ collisions.

Figure~\ref{story} (d) shows quantity $\nu$ (solid dots and curve) derived from product $x\nu$ inferred from \mmpt\ data and the \pbpb\ $\bar \rho_{sNN} \leftrightarrow x(n_s)/\alpha$ hypothesis shown in panel (a). 
Also shown are $\nu'$ values (open circles) derived from Glauber analysis  in Ref.~\cite{pbpbcent} that follow a power-law trend $\propto \bar \rho_0^{0.38}$ (dashed).
The \mmpt\ TCM data trend approaches the full Glauber power-law {\em variation} but {\em three times lower in magnitude}.

The trends in Fig.~\ref{story} are explained as follows: For peripheral collisions the three collision systems share a common \nn\ straight-line trend $x \approx \alpha \bar \rho_0$ with $\nu \approx 1$ corresponding to the congruences in Fig.~\ref{inverse} (left). \ppb\ and \pbpb\ collisions are restricted to single peripheral \nn\ collisions due to {\em exclusivity within a small overlap volume}~\cite{tomexclude}. \ppb\ and \pbpb\ trends then deviate from that linear trend above a transition near $\bar \rho_0 \approx \bar \rho_{s0} = 15$. The more-central trend for \ppb\ data can be accurately predicted because $N_{part} = N_{bin} + 1$ and all geometry parameters are therefore dependent on the single parameter $x(n_s)$ simply modeled as in Ref.~\cite{ppbpid}. The \pbpb\ geometry relies on developing a model for corresponding \mmpt\ data as in Fig.~\ref{inverse} and then following steps as in Fig.~\ref{story}.

\end{appendix}


\end{document}